\documentclass[twocolumn]{svjour3}
\usepackage{graphicx}
\usepackage{subfigure,graphicx}
\usepackage{amsmath}
\usepackage[]{placeins}
\usepackage{amsfonts}
\usepackage{float}
\usepackage{amssymb}
\usepackage{mathtools}
\usepackage{tensor}
\usepackage{enumitem}

\def\bfx{{\bf x}}
\def\bfy{{\bf y}}

\def\bfC{{\bf C}}

\def\bfI{{\bf I}}

\def\bfN{{\bf N}}

\def\bfS{{\bf S}}

\def\bfX{{\bf X}}
\def\bfF{{\bf F}}

\def\bfe{{\bf e}}

\def\Ktan{\mbox{\boldmath$\mathcal{K}$}}

\def\e0{\varepsilon_0}

\def\s0{\sigma_0}

\long\def\symbolfootnote[#1]#2{\begingroup%
\def\thefootnote{\fnsymbol{footnote}}\footnote[#1]{#2}\endgroup}

\long\def\symbolfootnote[#1]#2{\begingroup%
\def\thefootnote{\fnsymbol{footnote}}\footnote[#1]{#2}\endgroup}

\begin{document}

\journalname{International Journal of Fracture}
\titlerunning{The delayed fracture test for viscoelastic elastomers}

\title{The delayed fracture test for viscoelastic elastomers}

\author{B. Shrimali \and   O. Lopez-Pamies}

\institute{
           Bhavesh Shrimali \at Department of Civil and Environmental Engineering, University of Illinois, Urbana--Champaign, IL 61801-2352, USA\\
           \email{bshrima2@illinois.edu}\vspace{0.1cm}
           \and
           Oscar Lopez-Pamies \at
           Department of Civil and Environmental Engineering, University of Illinois, Urbana--Champaign, IL 61801-2352, USA  \\
           \email{pamies@illinois.edu}
           }

\date{Received: date / Accepted: date}
\maketitle

\begin{abstract}

In a recent contribution, Shrimali and Lopez-Pamies (2023) have shown that the Griffith criticality condition that governs crack growth in viscoelastic elastomers can be reduced --- from its ordinary form involving a historically elusive loading-history-dependent critical tearing energy $T_c$ --- to a fundamental form that involves exclusively the intrinsic fracture energy $G_c$ of the elastomer. The purpose of this paper is to make use of this fundamental form to explain one of the most distinctive fracture tests for viscoelastic elastomers, the so-called delayed fracture test.


\end{abstract}

\vspace{-0.5cm}

\section{Introduction} \label{Introduction}

It has been long established that the Griffith criticality condition
\begin{equation}
-\dfrac{\partial \mathcal{W}}{\partial \mathrm{\Gamma}_0}=T_{c}\label{Tc-0}
\end{equation}
describes the nucleation of fracture from pre-existing cracks --- as well as the propagation of fracture --- in elastomers subjected to quasi-static mechanical loads (Rivlin and Thomas, 1953; Greensmith and Thomas, 1955; Mullins, 1959; Lake and Thomas, 1967; Ahagon and Gent, 1975; Gent and Tobias, 1982; Gent, 1996; Tsunoda et al., 2000; Knauss, 2015).

The left-hand side $-\partial \mathcal{W}/\partial \mathrm{\Gamma}_0$ in expression (\ref{Tc-0}) stands for the change in total (stored and dissipated) deformation energy $\mathcal{W}$ in the elastomer with respect to an added surface area to the pre-existing crack $\mathrm{\Gamma}_0$ under conditions of fixed deformation of those parts of the boundary that are not traction-free so that no work is done by the external forces; note that the added surface area refers to the undeformed configuration.

The right-hand side $T_c$ is the so-called critical tearing energy. It is a characteristic property of the elastomer. Importantly, it is \emph{not} a constant. Much like $\mathcal{W}$, it is a function of the loading history. More specifically, experiments have established that $T_{c}$ can be written in the general form
\begin{equation*}
T_{c}=G_c(1+f_c).
\end{equation*}
In this expression, $G_c$ denotes the intrinsic fracture energy, or critical energy release rate, associated with the creation of new surface in the given elastomer. It is a material constant, independent of time. Its value is in the relatively narrow range $G_c\in[10,100]$ N/m for all common hydrocarbon elastomers (Ahagon and Gent, 1975; Gent and Tobias, 1982). On the other hand, $f_c$ is a non-negative function of the loading history that scales with the viscosity of the elastomer (Mullins, 1959; Gent and Lai, 1994; Knauss, 1973; Gent, 1996).  Precisely how $f_c$ --- and hence $T_c$ --- depends on the loading history for arbitrary loading conditions has remained an open problem for decades rendering the Griffith criticality condition in its ordinary form (\ref{Tc-0}) of limited practical utility.

In a recent contribution, Shrimali and Lopez-Pamies (2023) have uncovered a general formula for $f_c$ --- and hence $T_c$ --- and in so doing they have determined that (\ref{Tc-0}) can in fact be reduced to a form that involves \emph{not} the historically elusive critical tearing energy $T_c$, but only the intrinsic fracture energy $G_c$ of the elastomer. The result hinges on the following two elementary observations.
\begin{enumerate}

\item{For any viscoelastic elastomer, the total deformation energy $\mathcal{W}$ in (\ref{Tc-0}) can be partitioned into three different contributions:
\begin{equation}
\mathcal{W}=\underbrace{\mathcal{W}^{{\rm Eq}}+\mathcal{W}^{{\rm NEq}}}_\text{stored}+\underbrace{\mathcal{W}^{v}}_\text{dissipated}. \label{WWW}
\end{equation}
Here, $\mathcal{W}^{v}$ represents the part of the total energy that is dissipated by the elastomer via viscous deformation. On the other hand, the combination $\mathcal{W}^{{\rm Eq}}+\mathcal{W}^{{\rm NEq}}$ represents the part of the total energy that is stored by the elastomer via elastic deformation. In this combination, $\mathcal{W}^{{\rm NEq}}$ stands for the part of the stored elastic energy that will be dissipated eventually via viscous dissipation as the elastomer reaches a state of thermodynamic equilibrium. On the contrary, $\mathcal{W}^{{\rm Eq}}$ denotes the part of the stored elastic energy that the elastomer will retain at thermodynamic equilibrium.

\begin{figure}[h!]\centering
 \includegraphics[width=2.4in]{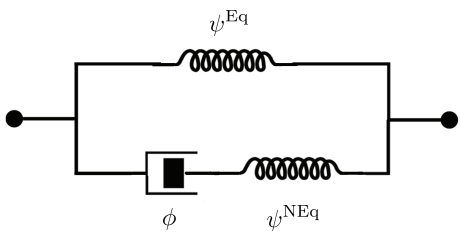}
\caption{\small A rheological representation of viscoelastic elastomers.}
   \label{Fig1}
\end{figure}
Rheological representations of elastomers provide a helpful visualization of the partition (\ref{WWW}). For instance, in the Zener-type rheological representation depicted in Fig. \ref{Fig1}, $\mathcal{W}^{{\rm Eq}}$ and $\mathcal{W}^{{\rm NEq}}$ correspond to the elastic energy stored in the equilibrium and non-equilibrium springs, where- as $\mathcal{W}^{v}$ corresponds to the viscous energy dissipated by the dashpot.
}

\item{``Pure-shear'' fracture tests of common hydrocarbon elastomers, as well as of more modern types of elastomers, consistently show --- rather remarkably --- that fracture occurs from the pre-existing crack in the specimens at a critical stretch that is independent (to within experimental error) of the stretch rate at which the test is carried out.}

\end{enumerate}
Precisely, by combining the above two observations, Shrimali and Lopez-Pamies (2023) have shown that the Griffith criticality condition (\ref{Tc-0}) can be reduced to the fundamental form
\begin{equation}
-\dfrac{\partial \mathcal{W}^{{\rm Eq}}}{\partial \mathrm{\Gamma}_0}=G_{c}.\label{Gc-0}
\end{equation}

From a physical point of view, the criticality condition (\ref{Gc-0}) states that whether an elastomer simply deforms or, on the other hand, creates new surface from a pre-existing crack is dictated by a competition between its stored equilibrium elastic energy and its intrinsic fracture energy, irrespective of its viscosity.

From a practical point of view, the criticality condition (\ref{Gc-0}) is straightforward to employ. This is because it is based on quantities that can be measured experimentally once and for all by means of conventional tests. Indeed, on the one hand, conventional experiments suffice to characterize the viscoelasticity of the elastomer of interest from which the storage of equilibrium elastic energy can then be identified; see, e.g., Section 4 in (Shrimali and Lopez-Pamies, 2023) and also Section \ref{Results Solithane} below. On the other hand, experiments in the spirit of those carried out, e.g., by  Gent and Tobias (1982) are enough to measure the intrinsic fracture energy $G_c$ of the elastomer.

What is more, as already noted above, the criticality condition (\ref{Gc-0}) brings resolution to the decades-old open problem of how the critical tearing energy $T_c$ depends on the loading history, as it entails that
\begin{equation}
T_c=G_c(1+ \,f_c)=G_c-\dfrac{\partial \mathcal{W}^{{\rm NEq}}}{\partial \mathrm{\Gamma}_0}-\dfrac{\partial \mathcal{W}^{v}}{\partial \mathrm{\Gamma}_0},\label{Tc-for}
\end{equation}
where the last two terms, $\partial \mathcal{W}^{{\rm NEq}}/\partial \mathrm{\Gamma}_0$ and $\partial \mathcal{W}^{v}/$ $\partial \mathrm{\Gamma}_0$, are to be evaluated at the instance in time at which the criticality condition (\ref{Gc-0}) is attained along the loading path of interest.

\begin{remark}\label{Remark0}

So as to provide a modicum of historical perspective, it is appropriate to make explicit mention of the various attempts at describing the critical tearing energy $T_c$ of elastomers that have been reported in the literature prior to the discovery of the general result (\ref{Tc-for}). A representative but non-exhaustive list includes the works of Knauss (1973), Schapery (1975;1984), Christensen (1979), de Gennes (1996), and Persson and Brener (2005). Invariably, all of these attempts are based on derivations that are centered around tearing (or peeling) experiments where a crack is \emph{propagated} at a constant velocity. Save for an exception (Schapery, 1984), all of them restrict attention to \emph{linear viscoelasticity}. Moreover, all of them make use either of a \emph{cohesive zone} or of an equivalent \emph{cutoff region} around the crack front, a constitutive assumption that further muddies their theoretical standing. By contrast, as already recalled above, the discovery of the general formula (\ref{Tc-for}) was made possible by centering its derivation around \emph{nucleation} of fracture in ``pure-shear'' fracture experiments, in particular, around the seemingly universal fact that fracture nucleation in such experiments takes places at critical stretches that are independent of the applied stretch rate. What is more, given its general status, the formula (\ref{Tc-for}) applies to any \emph{nonlinear viscoelastic} solid and is free of the constitutive restriction of having to explicitly identify a special region (the ``fracture process zone'') around the crack front.

\end{remark}

\begin{figure}[t!]
   \centering \includegraphics[width=3.in]{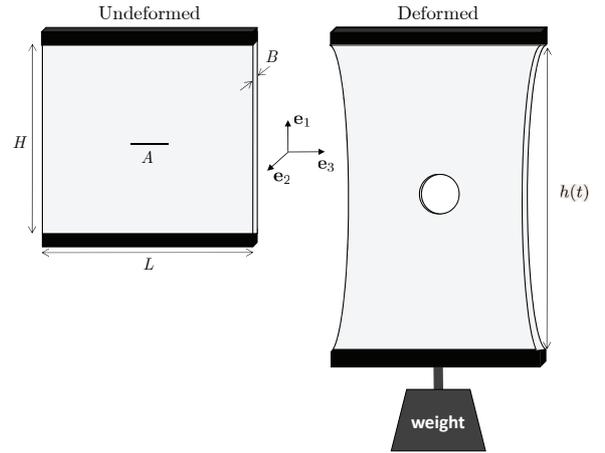}
   \caption{Schematic of a typical delayed fracture test for a viscoelastic elastomer. The specimen is held firmly by stiff grips. A load is applied rapidly from $t=0$ to $t=t_0\ll 1$ and then held constant. For a sufficiently large load, the nucleation of fracture from the pre-existing crack (of initial length $A$ here) may occur at a critical time $t_c>t_0$, hence the name of the test.}\label{Fig2}
\end{figure}

The object of this paper is to make use of the newly-minted fundamental form (\ref{Gc-0}) of the Griffith criticality condition in order to explain in a detailed and quantitative manner a tell-tale fracture test for viscoelastic elastomers: the so-called delayed fracture test. In a typical delayed fracture test, a sheet of the elastomer of interest containing a pre-exiting crack is subjected to a load that is applied rapidly over a very short time interval $[0,t_0\ll 1]$ and then held constant. Nucleation of fracture from the pre-existing crack occurs at a critical time $t_c>t_0$, hence the name of the test. In this work, consistent with the setup used by Knauss (1970) in his pioneering experiments, we will focus on the configuration depicted in Fig. \ref{Fig2}, where the pre-exiting crack is located in the center of the specimen and the load is applied in a uniaxial fashion.

The organization of the paper is as follows. We begin in Section \ref{Formulation} by formulating the pertinent initial-boundary-value problem. In Section \ref{Results Gaussian}, with the objective of exposing the chief characteristics of the delayed fracture test in the most basic of settings, we present and discuss sample generic results for the canonical case of a viscoelastic elastomer with Gaussian elasticity and constant viscosity. In Section \ref{Results Solithane}, we explain the experiments of Knauss (1970) on Solithane 113, a polyurethane elastomer with non-Gaussian elasticity and nonlinear viscosity. We conclude by recording a number of final comments in Section \ref{Sec: Final Comments}.

%
%

\section{Formulation of the initial-boundary-value problem for the delayed fracture test} \label{Formulation}

\subsection{Initial configuration and kinematics}

Consider the rectangular specimens depicted in Fig. \ref{Fig2} of length $L=101.6$ mm and height $H=L=101.6$ mm in the $\bfe_3$ and $\bfe_1$ directions and constant thickness $B=0.7938$ mm in the $\bfe_2$ direction. The specimens contain a pre-existing central crack of five different lengths
\begin{equation*}
A=2.5,5.08,10,15,20\; {\rm mm}
\end{equation*}
in the $\bfe_3$ direction. As alluded to above, these specific values for $L$, $H$, $B$, $A$ are chosen because they include those utilized by Knauss (1970) in his original delayed fracture experiments. Here, $\{\bfe_i\}$ stands for the laboratory frame of reference. We place its origin at the geometric center of the specimens so that, in their initial configuration at time $t=0$, the specimens occupy the domain
\begin{equation*}
\overline{\mathrm{\Omega}}_0=\{\bfX: \bfX\in\mathcal{P}_0\setminus\mathrm{\Gamma}_0\},
\end{equation*}
where
\begin{equation*}
\mathcal{P}_0=\left\{\bfX: |X_1|\leq\dfrac{H}{2},\,|X_2|\leq\dfrac{B}{2},\,   |X_3|\leq \dfrac{L}{2}  \right\}
\end{equation*}
and
\begin{equation*}
\mathrm{\Gamma}_0=\left\{\bfX: X_1=0,\,|X_2|\leq\dfrac{B}{2},\,  |X_3|\leq \dfrac{A}{2}  \right\}.
\end{equation*}

At a later time $t\in(0,T]$, due to the applied boundary conditions described below, the position vector $\bfX$ of a material point in the specimens will move to a new position specified by
\begin{equation*}
\bfx=\bfy(\bfX, t),
\end{equation*}
where $\bfy$ is a mapping from $\mathrm{\Omega}_0$ to the current configuration $\mathrm{\Omega}(t)$. We consider only invertible deformations, and write the deformation gradient field at $\bfX$ and $t$ as
\begin{equation*}
\bfF(\bfX, t)=\nabla\bfy(\bfX,t)=\frac{\partial \bfy}{\partial \bfX}(\bfX,t).
\end{equation*}

\subsection{Constitutive behavior of the elastomer}

The specimens are taken to be made of an isotropic incompressible elastomer. Making use of the two-potential formalism (Kumar and Lopez-Pamies, 2016), we describe its constitutive behavior by two thermodynamic potentials, the free energy
\begin{equation}
\psi(\bfF,\bfF^v)=\left\{\begin{array}{ll}\psi^{{\rm Eq}}(I_1)+\psi^{{\rm NEq}}(I_1^e)& \;{\rm if}\; J=1\\ \\
+\infty & {\rm otherwise}\end{array}\right.\label{psi}
\end{equation}
that describes how the elastomer stores energy through elastic deformation and the dissipation potential
\begin{equation}
\phi(\bfF,\bfF^v,\dot{\bfF}^v)=\left\{\begin{array}{ll}
\dfrac{1}{2}\dot{\bfF}^v{\bfF^v}^{-1}\cdot\left[2\,\eta(I_1^e,I_2^e,I_1^v)\times\right. &\\
\left.\Ktan\,\dot{\bfF}^v{\bfF^v}^{-1}\right] & \hspace{-2cm} {\rm if}\;{\rm tr}( \dot{\bfF}^v{\bfF^v}^{-1})=0 \vspace{0.2cm} \\
+\infty & \hspace{-1cm} {\rm otherwise}\end{array}\right. \label{phi}
\end{equation}
that describes how the elastomer dissipates energy through viscous deformation. In these expressions, the second-order tensor $\bfF^v$ is an internal variable of state that describes roughly the ``viscous part'' of the deformation gradient $\bfF$, the ``dot'' notation stands for the Lagrangian time derivative (i.e., with $\bfX$ held fixed),
\begin{align*}
&I_1={\rm tr}\,\bfC, \quad  J=\sqrt{\det\bfC},\\
& I^v_1={\rm tr}\,\bfC^v, \quad I_1^e={\rm tr}(\bfC{\bfC^{v}}^{-1}),\\ & I_2^e=\dfrac{1}{2}\left[\left(\bfC\cdot{\bfC^v}^{-1}\right)^2-{\bfC^v}^{-1}\bfC\cdot\bfC{\bfC^v}^{-1}\right],
\end{align*}
where $\bfC=\bfF^T\bfF$ denotes the right Cauchy-Green deformation tensor, $\bfC^v={\bfF^v}^T\bfF^v$, $\mathcal{K}_{ijkl}=\frac{1}{2}(\delta_{ik}\delta_{jl}+\delta_{il}\delta_{jk}-\frac{2}{3}\delta_{ij}\delta_{kl})$ stands for the standard deviatoric orthogonal projection tensor, and $\psi^{{\rm Eq}}$, $\psi^{{\rm NEq}}$, $\eta$ are any (suitably well-behaved) non-negative material functions of their arguments.

Granted the two thermodynamic potentials (\ref{psi}) and (\ref{phi}), it follows that the first Piola-Kirchhoff stress tensor $\bfS$ at any material point $\bfX\in\mathrm{\Omega}_0$ and time $t\in[0,T]$ is expediently given by the relation (Kumar and Lopez-Pamies, 2016)
\begin{equation*}
\bfS(\bfX,t)=\frac{\partial \psi}{\partial\bfF}(\bfF,\bfF^v),
\end{equation*}
where $\bfF^v$ is implicitly defined by the evolution equation
\begin{equation*}
\dfrac{\partial \psi}{\partial \bfF^v}(\bfF,\bfF^v)+\dfrac{\partial \phi}{\partial \dot{\bfF}^v}(\bfF,\bfF^v,\dot{\bfF}^v)={\bf0}.
\end{equation*}
%
%
%
Making use of the specific isotropic incompressible forms (\ref{psi}) and (\ref{phi}), this relation can be rewritten more explicitly as
\begin{equation}
\bfS(\bfX,t)=2\psi^{{\rm Eq}}_{I_1}\bfF+2\psi^{{\rm NEq}}_{I^e_1}\bfF{\bfC^v}^{-1}-p\bfF^{-T},\label{S-I1-J}
\end{equation}
where $p$ stands for the arbitrary hydrostatic pressure associated with the incompressibility constraint $J=1$ of the elastomer, $\bfC^v$ is defined implicitly as the solution of the evolution equation
\begin{equation}
\dot{\bfC}^v(\bfX,t)=\dfrac{2\psi^{{\rm NEq}}_{I^e_1}}{\eta(I_1^e,I_2^e,I_1^v)}\left[\bfC-\dfrac{1}{3}\left(\bfC\cdot{\bfC^v}^{-1}\right)\bfC^v\right],  \label{Evolution-I1-J}
\end{equation}
%
%
%
and where we have made use of the notation $\psi^{{\rm Eq}}_{I_1}={\rm d} \psi^{{\rm Eq}}(I_1)/{\rm d}  I_1$ and $\psi^{{\rm NEq}}_{I^e_1}={\rm d} \psi^{{\rm NEq}}$ $(I^e_1)/{\rm d}  I^e_1$. Note that the dependence on the internal variable $\bfF^v$ ends up entering (\ref{S-I1-J}) and (\ref{Evolution-I1-J}) only through the symmetric combination $\bfC^v={\bfF^v}^T\bfF^v$.

For a detailed account of the constitutive relation (\ref{S-I1-J})-(\ref{Evolution-I1-J}), the interested reader is referred to (Kumar and Lopez-Pamies, 2016). Here, we remark that the constitutive relation (\ref{S-I1-J})-(\ref{Evolution-I1-J}) corresponds to a generalization of the classical Zener or standard solid model (Zener, 1948) to the setting of  finite deformations. Accordingly, as schematically depicted by the rheological representation in Fig. \ref{Fig1}, the function $\psi^{{\rm Eq}}$ in (\ref{psi}) characterizes the elastic energy storage in the elastomer at states of thermodynamic equilibrium, whereas $\psi^{{\rm NEq}}$ characterizes the additional elastic energy storage at non-equilibrium states (i.e., again, the part of the energy that gets dissipated eventually). On the other hand, the function $\eta$ in (\ref{phi}) characterizes the  viscosity of the elastomer.

%
%
%
%
%
%

In the results that are presented in Sections \ref{Results Gaussian} and \ref{Results Solithane} below, we will make use of the following specific forms for the equilibrium and non-equilibrium free-energy functions in (\ref{psi}) and viscosity function in (\ref{phi}):
\begin{equation}
\left\{\hspace{-0.1cm}\begin{array}{l}
\psi^{{\rm Eq}}(I_1)=\displaystyle\sum\limits_{r=1}^N\dfrac{3^{1-\alpha_r}}{2 \alpha_r}\mu_r\left(\,I_1^{\alpha_r}-3^{\alpha_r}\right)\vspace{0.2cm}\\
\psi^{{\rm NEq}}(I^e_1)=\displaystyle\sum\limits_{r=1}^N\dfrac{3^{1-\beta_r}}{2 \beta_r}\nu_r\left(\,{I^e_1}^{\beta_r}-3^{\beta_r}\right)\vspace{0.2cm}\\
\eta(I_1^e,I_2^e,I_1^v)=\eta_{\infty}+\dfrac{\eta_0-\eta_{\infty}+K_1\left[{I_1^v}^{\gamma_1}-3^{\gamma_1}\right]}{1+\left(K_2 \mathcal{J}_2^{{\rm NEq}}\right)^{\gamma_2}}
\end{array}\right. \label{Prescriptions}
\end{equation}
with $\mathcal{J}_2^{{\rm NEq}}=({I_1^e}^2/3-I_2^e)(\sum_{r=1}^N 3^{1-\beta_r}\nu_r {I_1^e}^{\beta_r-1})^2$ and $N=1,2$, which result in the constitutive relation
\begin{align}
\bfS(\bfX,t)=&\displaystyle\sum\limits_{r=1}^N 3^{1-\alpha_r} \mu_r I_1^{\alpha_r-1} \bfF+\nonumber\\
&\displaystyle\sum\limits_{r=1}^N 3^{1-\beta_r}\nu_r {I^e_1}^{\beta_r-1}\bfF{\bfC^v}^{-1}-p\bfF^{-T}\label{S-KLP}
\end{align}
with evolution equation
\begin{align}
\dot{\bfC}^v(\bfX,t)=&\dfrac{\sum\limits_{r=1}^N 3^{1-\beta_r}\nu_r {I^e_1}^{\beta_r-1}}{\eta_{\infty}+\frac{\eta_0-\eta_{\infty}+K_1\left[{I_1^v}^{\gamma_1}-3^{\gamma_1}\right]}{1+\left(K_2 \mathcal{J}_2^{{\rm NEq}}\right)^{\gamma_2}}}\left[\bfC-\dfrac{1}{3}\times\right.\nonumber\\
&\left.\left(\bfC\cdot{\bfC^v}^{-1}\right)\bfC^v\right]. \label{Evolution-KLP}
\end{align}

The constitutive prescription (\ref{S-KLP})-(\ref{Evolution-KLP}) includes several fundamental constitutive relations as special cases. For instance, it includes the case of a Neo-Hookean solid ($N=1$, $\nu_1=0$, $\alpha_1=1$, $\eta_0=\eta_{\infty}=0$, $K_1=K_2=0$), that of a Newtonian fluid ($N=1$, $\mu_1=0$, $\nu_1=+\infty$, $\eta_{\infty}=0$, $K_1=K_2=0$), as well as that of a viscoelastic elastomer with Gaussian elasticity and constant viscosity ($N=1$, $\alpha_1=\beta_1=1$, $\eta_{\infty}=0$, $K_1=K_2=0$). What is more, the prescription (\ref{S-KLP})-(\ref{Evolution-KLP}) has been shown to be accurately descriptive and predictive of a wide range of elastomers, which typically exhibit non-Gaussian elasticity as well as nonlinear viscosity of shear-thinning type (Lopez-Pamies, 2010; Kumar and Lopez-Pamies, 2016; Ghosh and Lopez-Pamies, 2021; Chockalingam et al., 2021, Chen and Ravi-Chandar, 2022). In all, note that the constitutive prescription (\ref{S-KLP})-(\ref{Evolution-KLP}) contains $4N+6$ material parameters. $2N$ of them, $\mu_r$ and $\alpha_r$ ($r=1,...,N$), serve to characterize the non-Gaussian elasticity of the elastomer at states of thermodynamic equilibrium. Another $2N$, $\nu_r$ and $\beta_r$ ($r=1,...,N$), characterize the non-Gaussian elasticity  at non-equilibrium states. Finally, the last six parameters, $\eta_0$, $\eta_{\infty}$, $K_1$, $K_2$, $\gamma_1$, $\gamma_2$, serve to characterize the nonlinear shear-thinning viscosity.

\subsection{Initial and boundary conditions}

In their initial configuration, we consider that the specimens are undeformed and stress-free. Therefore, we have the initial conditions
\begin{equation}
\left\{\begin{array}{l}
\bfy(\bfX,0)=\bfX\vspace{0.2cm}\\
p(\bfX,0)=2\psi^{{\rm Eq}}_{I_1}(3)+2\psi^{{\rm NEq}}_{I^e_1}(3)\vspace{0.2cm}\\
\bfC^v(\bfX,0)=\bfI \end{array}\right., \quad\bfX\in \overline{\mathrm{\Omega}}_0.\label{ICs}
\end{equation}

The top
\begin{equation*}
\partial\mathrm{\Omega}^{\mathcal{T}}_0=\left\{\bfX: X_1=\dfrac{H}{2},\,|X_2|\leq\dfrac{B}{2},\,  |X_3|\leq \dfrac{L}{2}  \right\}
\end{equation*}
and the bottom boundary
\begin{equation*}
\partial\mathrm{\Omega}^{\mathcal{B}}_0=\left\{\bfX: X_1=-\dfrac{H}{2},\,|X_2|\leq\dfrac{B}{2},\,  |X_3|\leq \dfrac{L}{2}  \right\}
\end{equation*}
of the specimens are held firmly by stiff grips on which a force of magnitude
\begin{equation}
P(t)=\left\{\begin{array}{ll}
\dfrac{2\sigma_0(B\times L) t_0 t}{t_0^2+t^2}  & \qquad {\rm if} \quad 0\leq t\leq t_0 \vspace{0.15cm}\\
\sigma_0(B\times L)  & \qquad {\rm if} \quad t_0<t\leq T
\end{array}\right.\label{P(t)}
\end{equation}
is applied in the $\pm\bfe_1$ directions resulting in a separation $h(t)$ between the grips; see Fig. \ref{Fig2}. In the results that are presented in Sections \ref{Results Gaussian} and \ref{Results Solithane}, consistent, once more, with the experiments of Knauss (1970), we make use of the values
\begin{equation}
t_0=0.01\, {\rm s}\quad {\rm and}\quad \sigma_0\in[0,0.3]\, {\rm MPa},\label{t0}
\end{equation}
which correspond to a force $P_0=\sigma_0(B\times L)$ that is applied rapidly over the very short time interval $[0,t_0]$ and then held constant. The rest of the boundary $\partial\mathrm{\Omega}_0$ of the specimens is traction free. Precisely, making use of the notation $\textbf{s}(\bfX,t)=\bfS\bfN$, we have that
\begin{equation}
\left\{\hspace{-0.05cm}\begin{array}{ll}
y_1(\bfX,t)=\dfrac{h(t)}{2}, & (\bfX,t)\in\partial\mathrm{\Omega}^{\mathcal{T}}_0\times[0,T] \vspace{0.15cm}\\
y_3(\bfX,t)=X_3, & (\bfX,t)\in\partial\mathrm{\Omega}^{\mathcal{T}}_0\times[0,T] \vspace{0.15cm}\\
\int_{\partial\mathrm{\Omega}^{\mathcal{T}}_0}s_1(\bfX,t){\rm d}\bfX=P(t), & (\bfX,t)\in\partial\mathrm{\Omega}^{\mathcal{T}}_0\times[0,T] \vspace{0.15cm}\\
s_2(\bfX,t)=0, & (\bfX,t)\in\partial\mathrm{\Omega}^{\mathcal{T}}_0\times[0,T] \vspace{0.15cm}\\
y_1(\bfX,t)=-\dfrac{h(t)}{2}, & (\bfX,t)\in\partial\mathrm{\Omega}^{\mathcal{B}}_0\times[0,T] \vspace{0.15cm}\\
y_3(\bfX,t)=X_3, & (\bfX,t)\in\partial\mathrm{\Omega}^{\mathcal{B}}_0\times[0,T] \vspace{0.15cm}\\
\int_{\partial\mathrm{\Omega}^{\mathcal{B}}_0}s_1(\bfX,t){\rm d}\bfX=-P(t), & (\bfX,t)\in\partial\mathrm{\Omega}^{\mathcal{B}}_0\times[0,T] \vspace{0.15cm}\\
s_2(\bfX,t)=0, & (\bfX,t)\in\partial\mathrm{\Omega}^{\mathcal{B}}_0\times[0,T] \vspace{0.15cm}\\
\textbf{s}=\textbf{0}, & \hspace{-2.35cm}(\bfX,t)\in\partial\mathrm{\Omega}_0\setminus\left(\partial\mathrm{\Omega}^{\mathcal{T}}_0\cup\partial\mathrm{\Omega}^{\mathcal{B}}_0\right)\times[0,T]
\end{array}\right. ,\label{BCs}
\end{equation}
where $\bfN$ stands for the outward unit normal to the boundary $\partial\mathrm{\Omega}_0$.

\begin{remark}\label{Remark1} In experiments, specimens like the ones of interest here are typically gripped in a way that complex triaxial stresses develop near the grips. Numerical experiments indicate that these localized stresses have practically no effect on the response of the specimens, thus our idealized choice of zero traction (\ref{BCs})$_{4,8}$ at the top and bottom boundaries.
\end{remark}

\begin{remark}\label{Remark2} In all the numerical solutions that are presented below, the mixed boundary conditions (\ref{BCs})$_{1,5}$ with (\ref{BCs})$_{3,7}$ are enforced by modeling explicitly the grips holding the specimens as nonlinear elastic materials with a stiffness 6 orders of magnitude larger than the elastomer being tested; see Fig. \ref{Fig5}.
\end{remark}

\subsection{Governing equations}

Upon putting all the above ingredients together, neglecting inertia and body forces, the mechanical response of the specimens is governed by the equilibrium and incompressibility constraint equations
\begin{equation}
\left\{\begin{array}{ll}{\rm Div}\,\bfS={\bf0}, & \quad (\bfX,t)\in\mathrm{\Omega}_0\times[0,T] \vspace{0.2cm} \\
\det\nabla\bfy=1, & \quad (\bfX,t)\in\mathrm{\Omega}_0\times[0,T]
\end{array}\right. \label{Equilibrium-PDE}
\end{equation}
subject to the initial and boundary conditions (\ref{ICs})$_{1,2}$ and (\ref{BCs}), where $\bfS(\bfX,t)=2\psi^{{\rm Eq}}_{I_1}\nabla\bfy+2\psi^{{\rm NEq}}_{I^e_1}\nabla\bfy{\bfC^v}^{-1}-p\nabla\bfy^{-T}$, coupled with the evolution equation
\begin{align}
\dot{\bfC}^v=\dfrac{2\psi^{{\rm NEq}}_{I^e_1}}{\eta(I_1^e,I_2^e,I_1^v)}\left[\nabla\bfy^T\nabla\bfy-\dfrac{1}{3}\left(\nabla\bfy^T\nabla\bfy\cdot{\bfC^v}^{-1}\right)\bfC^v\right], \label{Evolution-ODE}
\end{align}
subject to the initial condition (\ref{ICs})$_3$, for the deformation field $\bfy(\bfX,t)$, the pressure field $p(\bfX,t)$, and the internal variable $\bfC^v(\bfX,t)$.

In the next two sections, we present numerical solutions for the initial-boundary-value problem (\ref{Equilibrium-PDE})-(\ref{Evolution-ODE}) with (\ref{ICs})-(\ref{BCs}) and (\ref{Prescriptions}) for two sets of material parameters. First, in Section \ref{Results Gaussian}, we generate results for the canonical case of an elastomer with Gaussian elasticity and constant viscosity. In Section \ref{Results Solithane}, we generate results for the polyurethane elastomer studied by Knauss (1970). All the results that we present in the sequel are generated by a plane-stress variant of the numerical scheme introduced by Ghosh et al. (2021), which is based on a Crouzeix-Raviart finite-element discretization of space and a high-order explicit Runge-Kutta discretization of time.

\subsection{The computation of the energy release rate $-\partial \mathcal{W}^{{\rm Eq}}/\partial \mathrm{\Gamma}_0$ under boundary conditions of traction} \label{Griffith General}

In the present setting, the equilibrium elastic energy $\mathcal{W}^{{\rm Eq}}$ in the Griffith criticality condition (\ref{Gc-0}) is given by
\begin{align}
\mathcal{W}^{{\rm Eq}}(h(t),\mathrm{\Gamma}_0)=\displaystyle\int_{\mathrm{\Omega}_0}\psi^{{\rm Eq}}(I_1)\,{\rm d}\bfX,\label{WEq-u-G}
\end{align}
where we have made explicit the fact that $\mathcal{W}^{{\rm Eq}}$ can be thought of as a function of the deformation history $h(t)$ between the grips and the initial surface area $\mathrm{\Gamma}_0$ of the pre-existing crack.

In the present problem, however, the deformation history $h(t)$ between the grips is not prescribed and hence it is not known explicitly. It is only known implicitly in terms of the applied force $P(t)$ and the solution of the initial-boundary-value problem (\ref{Equilibrium-PDE})-(\ref{Evolution-ODE}) with (\ref{ICs})-(\ref{BCs}) and (\ref{Prescriptions}) for a given $\mathrm{\Gamma}_0$. With some abuse of notation, we write
\begin{align}
\mathcal{W}^{{\rm Eq}}(h(t),\mathrm{\Gamma}_0)=\mathcal{W}^{{\rm Eq}}(h(P(t),\mathrm{\Gamma}_0),\mathrm{\Gamma}_0).\label{WEqWEq}
\end{align}
%

%
%
%



By definition, the derivative $-\partial \mathcal{W}^{{\rm Eq}}/\partial \mathrm{\Gamma}_0$ in (\ref{Gc-0}) is to be carried out at fixed $h(t)$. In view of the arguments in the functional description (\ref{WEqWEq}) of the equilibrium elastic energy, this can be accomplished as follows.

Given a specimen with initial surface area $\mathrm{\Gamma}_0$ of the pre-existing crack and given an applied force $P(t)$, consider the addition of an increment ${\rm d}\mathrm{\Gamma}_0$ to $\mathrm{\Gamma}_0$, this at fixed $P(t)$. On use of the condition ${\rm d} P=0$, the associated incremental change in the equilibrium elastic energy $\mathcal{W}^{{\rm Eq}}$ reads
\begin{align*}
{\rm d} \mathcal{W}^{{\rm Eq}}=&\dfrac{\partial \mathcal{W}^{{\rm Eq}}}{\partial h}\left[\dfrac{\partial h}{\partial P}{\rm d} P+\dfrac{\partial h}{\partial \mathrm{\Gamma}_0}{\rm d} \mathrm{\Gamma}_0\right]+\dfrac{\partial \mathcal{W}^{{\rm Eq}}}{\partial \mathrm{\Gamma}_0}{\rm d} \mathrm{\Gamma}_0\\
=&\dfrac{\partial \mathcal{W}^{{\rm Eq}}}{\partial h}\dfrac{\partial h}{\partial \mathrm{\Gamma}_0}{\rm d} \mathrm{\Gamma}_0+\dfrac{\partial \mathcal{W}^{{\rm Eq}}}{\partial \mathrm{\Gamma}_0}{\rm d} \mathrm{\Gamma}_0.
\end{align*}
After a simple algebraic manipulation, it follows that
\begin{align}
-\dfrac{\partial \mathcal{W}^{{\rm Eq}}}{\partial \mathrm{\Gamma}_0}{\rm d} \mathrm{\Gamma}_0=& P^{\rm Eq}{\rm d} h-{\rm d} \mathcal{W}^{{\rm Eq}}\nonumber\\
=& {\rm d}\mathcal{W}^{{\rm Eq}\ast}-[h-H]{\rm d}P^{\rm Eq},\label{partialWstar}
\end{align}
where we have made use of the relation ${\rm d}h=(\partial h/$ $\partial\mathrm{\Gamma}_0)$ ${\rm d}\mathrm{\Gamma}_0$ and, for convenience, have introduced the notation
\begin{align}
P^{\rm Eq}(P(t),\mathrm{\Gamma}_0):=\dfrac{\partial \mathcal{W}^{{\rm Eq}}}{\partial h}\left(h(P(t),\mathrm{\Gamma}_0),\mathrm{\Gamma}_0\right)\label{PEquilibrium}
\end{align}
and
\begin{align}
\mathcal{W}^{{\rm Eq}\ast}(P(t),\mathrm{\Gamma}_0):= & P^{\rm Eq}(P(t),\mathrm{\Gamma}_0)\left[h(P(t),\mathrm{\Gamma}_0)-H\right]\nonumber\\
&-\mathcal{W}^{{\rm Eq}}\left(h(P(t),\mathrm{\Gamma}_0),\mathrm{\Gamma}_0\right).\label{Wstar}
\end{align}
It follows immediately from (\ref{partialWstar}) that
\begin{align}
-\dfrac{\partial \mathcal{W}^{{\rm Eq}}}{\partial \mathrm{\Gamma}_0}=&\dfrac{\partial \mathcal{W}^{{\rm Eq}\ast}}{\partial\mathrm{\Gamma}_0}(P(t),\mathrm{\Gamma}_0)-\nonumber\\
&[h(P(t),\mathrm{\Gamma}_0) -H]\dfrac{\partial P^{\rm Eq}}{\partial\mathrm{\Gamma}_0}(P(t),\mathrm{\Gamma}_0),\label{ERRP}
\end{align}
which is precisely the result that we are after. Indeed, given the applied force (\ref{P(t)}) in the delayed fracture tests of interest here, the result (\ref{ERRP}) allows us to expediently determine the resulting energy release rate $-\partial \mathcal{W}^{{\rm Eq}}/\partial \mathrm{\Gamma}_0$ in the Griffith criticality condition (\ref{Gc-0}) in terms of three readily computable quantities: the deformation $h(t)$ between the grips and the derivatives with respect to $\mathrm{\Gamma}_0$ at fixed $P(t)$ --- or, equivalently, at fixed time $t$ --- of the equilibrium elastic force (\ref{PEquilibrium}) and the complementary equilibrium elastic energy (\ref{Wstar}).

\section{Results for a canonical elastomer with Gaussian elasticity and constant viscosity}\label{Results Gaussian}

In this section, we present solutions for the initial-boundary-value problem (\ref{Equilibrium-PDE})-(\ref{Evolution-ODE}) with (\ref{ICs})-(\ref{BCs}) and (\ref{Prescriptions}) for the basic case when the specimen is made of a canonical elastomer with Gaussian elasticity and constant viscosity. Specifically, we present solutions for the case when $N=1$, $\alpha_1=\beta_1=1$, $\eta_{\infty}=0$, $K_1=K_2=0$, equilibrium and non-equilibrium initial shear moduli
\begin{equation*}
\mu_1=0.2\;{\rm MPa}\quad {\rm and}\quad \nu_1=2\; {\rm MPa},
\end{equation*}
and viscosity
\begin{equation*}
\eta_0=500\; {\rm MPa}\,{\rm s}.
\end{equation*}
These values are chosen here because they are comparable with those that describe the elastomer analyzed in the next section; see Table \ref{Table1}. Note that these material parameters correspond to an elastomer with constant relaxation time $\tau=\eta_0 \nu_1^{-1}=250$ ${\rm s}$ and constant creep time $\tau^\ast=\eta_0(\mu_1^{-1}+\nu_1^{-1})=2750$ ${\rm s}$.

\subsection{The force-deformation and deformation-time responses}

%
\begin{figure}[b!]
   \centering \includegraphics[width=2.6in]{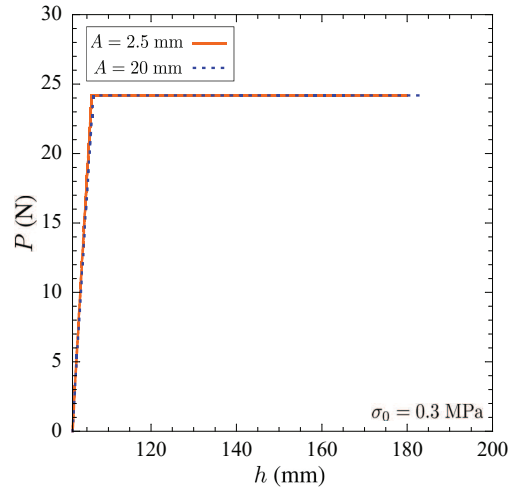}
   \caption{Force-deformation response of specimens with pre-existing cracks of lengths $A=2.5$ and $20$ mm for the applied force (\ref{P(t)})-(\ref{t0})$_1$ with global stress $\sigma_0=0.3$ MPa and total time of applied loading $T=20000$ s.}\label{Fig3}
\end{figure}
%
%
\begin{figure}[t!]
   \centering \includegraphics[width=2.6in]{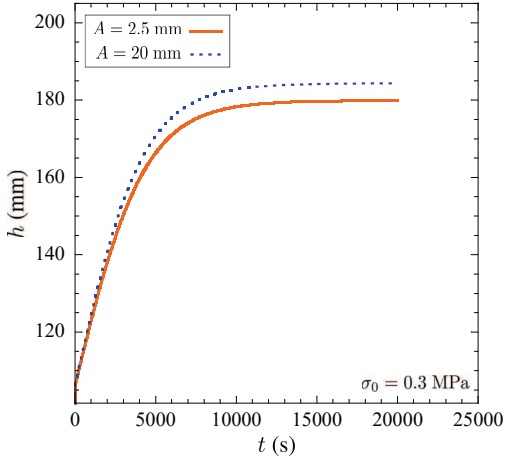}
   \caption{Evolution in time $t$ of the deformation $h(t)$ between the grips in specimens with pre-existing cracks of lengths $A=2.5$ and $20$ mm subjected to the applied force (\ref{P(t)})-(\ref{t0})$_1$ with global stress $\sigma_0=0.3$ MPa and total time of applied loading $T=20000$ s.}\label{Fig4}
\end{figure}
%
%
\begin{figure}[h!]
   \centering \includegraphics[width=3.in]{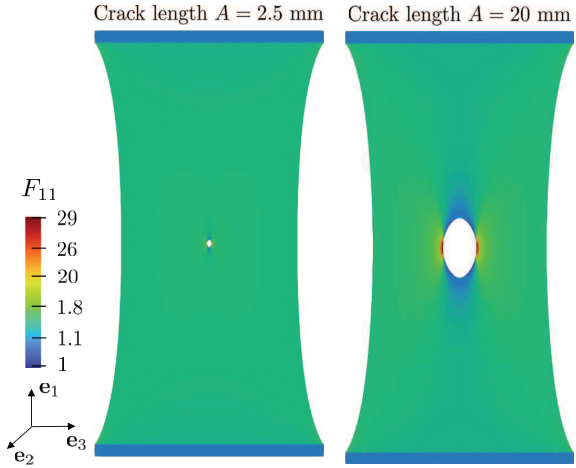}
   \caption{Contour plots over the deformed configuration of the component $F_{11}(\bfX,t)$ of the local deformation gradient in specimens with pre-existing cracks of lengths $A=2.5$ and $20$ mm subjected to the applied force (\ref{P(t)})-(\ref{t0})$_1$ with global stress $\sigma_0=0.3$ MPa. Both plots are shown at the same final time $t=T=20000$ s of the applied load.}\label{Fig5}
\end{figure}
%

Figures \ref{Fig3} and \ref{Fig4} present solutions for the deformation $h(t)$ between the grips that results from the applied force (\ref{P(t)})-(\ref{t0})$_1$ with global stress $\sigma_0=0.3$ MPa and total time of applied loading $T=20000$ s in specimens with pre-existing cracks of lengths $A=2.5$ and $20$ mm. Specifically, the results are shown for the applied force $P(t)$ as a function of $h(t)$ in Fig. \ref{Fig3} and for the evolution of $h(t)$ in time $t$ in Fig. \ref{Fig4}. To aid in the visualization of the results, Fig. \ref{Fig5} also shows contour plots over the deformed configuration of the component $F_{11}(\bfX,t)$ of the local deformation gradient at the same final time $t=T=20000$ s of the applied load for both specimens.

As expected, the specimen with the larger crack leads to a larger deformation between the grips for the same applied force. It is also interesting to note that by approximately $t=10000$ s --- at which point $h\approx 178$ mm in the specimen with crack length $A=2.5$ mm and $h\approx 183$ mm in that with $A=20$ mm --- the creeping process has all but concluded, this for both specimens. Finally, we remark that the results for other values of global stress $\sigma_0$ in the range (\ref{t0})$_2$ are not fundamentally different from those shown in Figs. \ref{Fig3} through \ref{Fig5} for $\sigma_0=0.3$ MPa.

\subsection{The total deformation energy $\mathcal{W}$ and its partition into $\mathcal{W}^{{\rm Eq}}$, $\mathcal{W}^{{\rm NEq}}$, and $\mathcal{W}^{v}$}

%
\begin{figure}[t!]
  \subfigure[]{
   \label{fig:7a}
   \begin{minipage}[]{0.5\textwidth}
   \centering \includegraphics[width=2.75in]{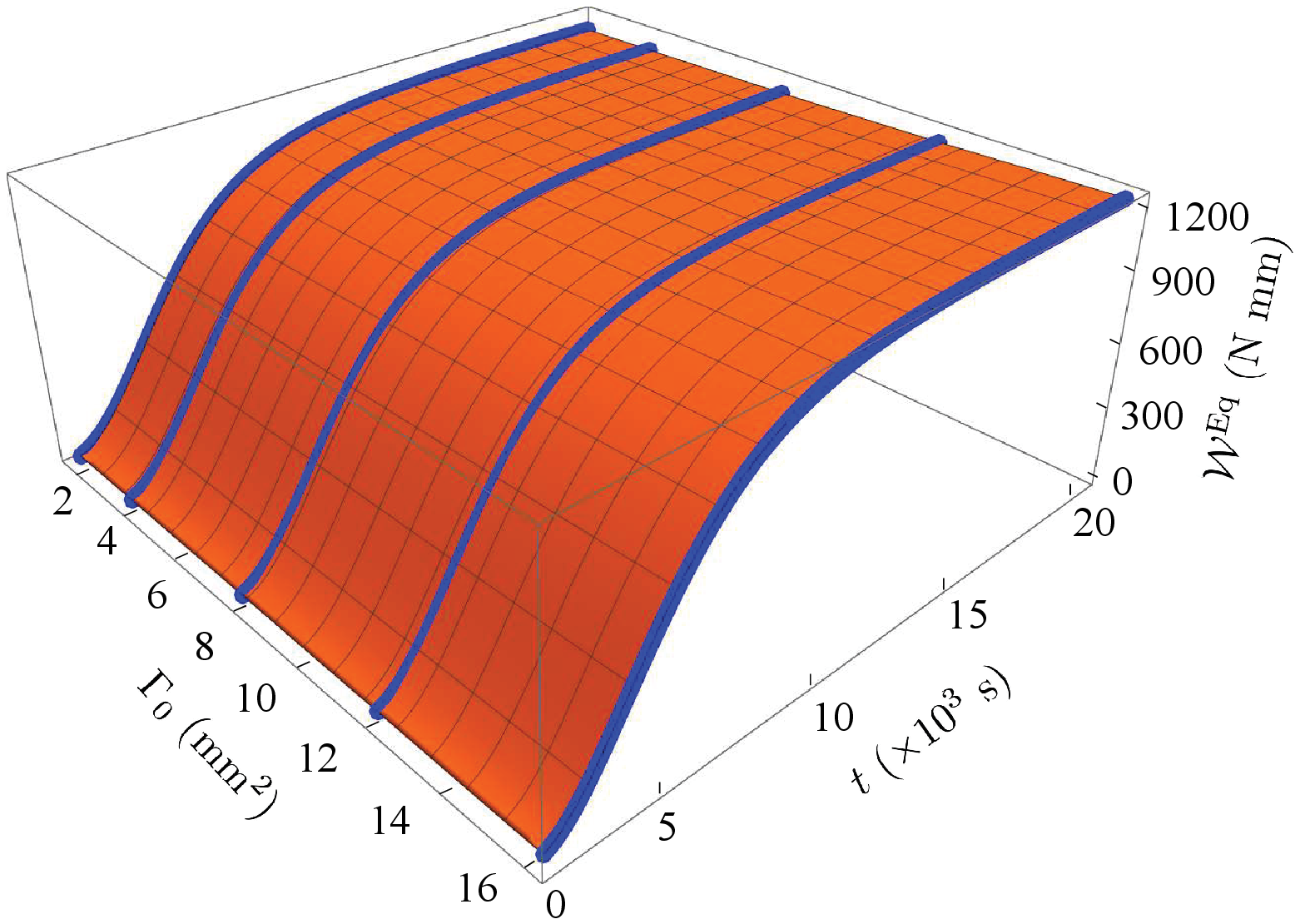}
   \vspace{0.2cm}
   \end{minipage}}
  \subfigure[]{
   \label{fig:7b}
   \begin{minipage}[]{0.5\textwidth}
   \centering \includegraphics[width=2.75in]{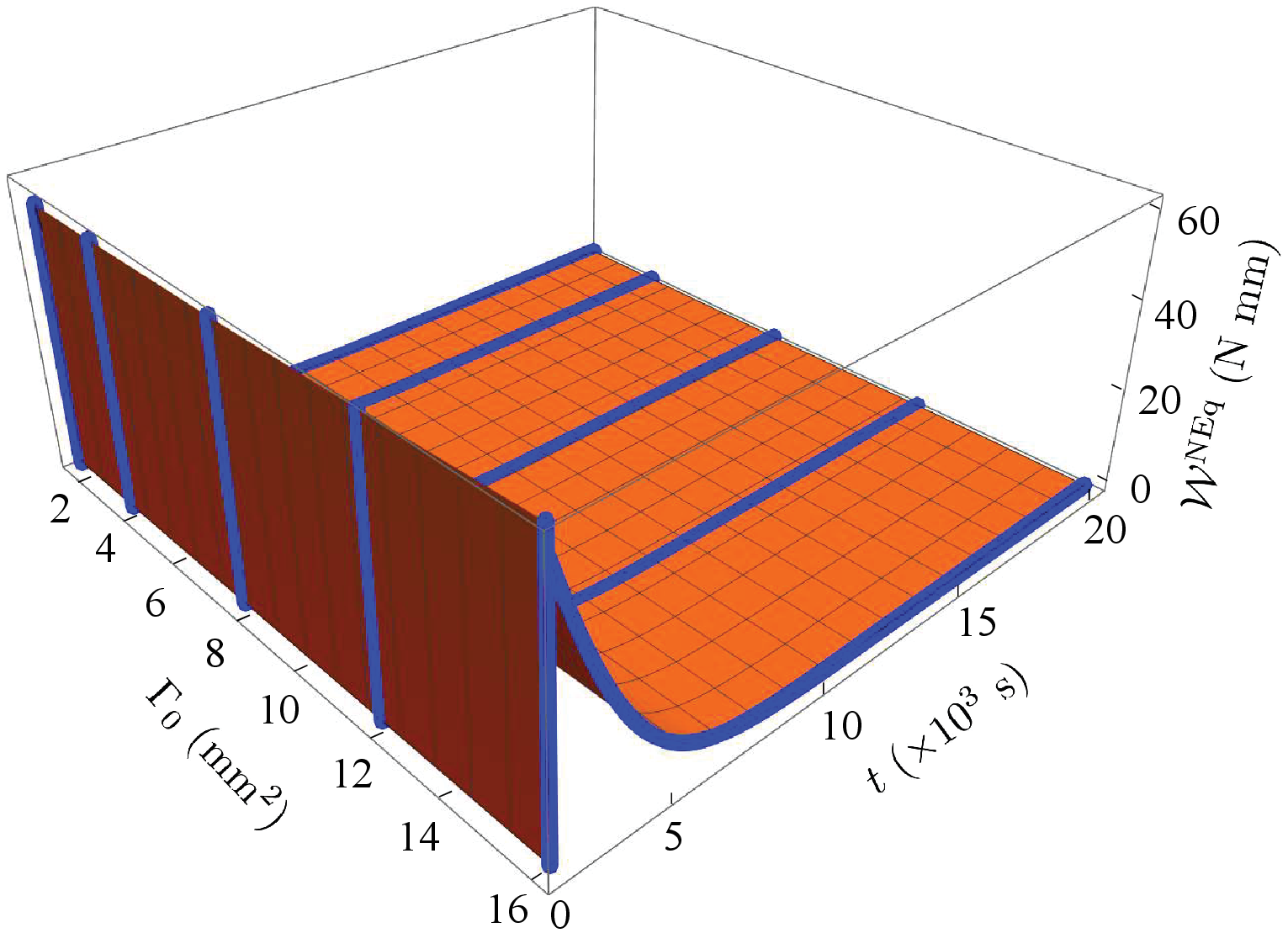}
   \vspace{0.2cm}
   \end{minipage}}
     \subfigure[]{
   \label{fig:7c}
   \begin{minipage}[]{0.5\textwidth}
   \centering \includegraphics[width=2.75in]{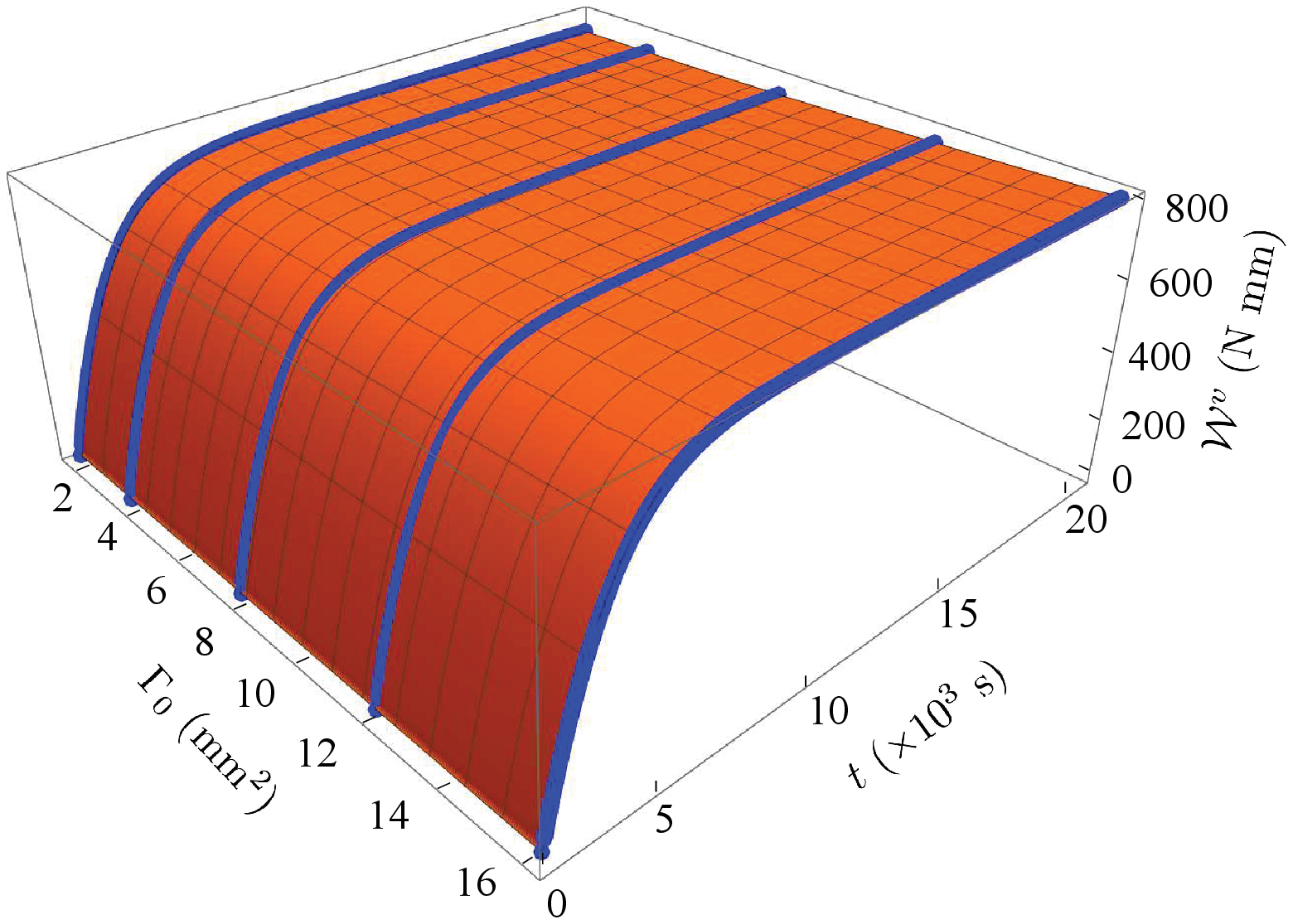}
   \vspace{0.2cm}
   \end{minipage}}
   \caption{Computed values from (\ref{WEq-NH})-(\ref{Wv-NH}) of (a) the equilibrium elastic energy $\mathcal{W}^{{\rm Eq}}$, (b) the non-equilibrium elastic energy $\mathcal{W}^{{\rm NEq}}$, and (c) the dissipated viscous energy $\mathcal{W}^{v}$ in specimens subjected to the applied force (\ref{P(t)})-(\ref{t0})$_1$, with global stress $\sigma_0=0.3$ MPa and total time of applied loading $T=20000$ s, plotted as functions of the initial crack surface $\mathrm{\Gamma}_0=A \times B$ and time $t$.}\label{Fig6}
\end{figure}
%

%
\begin{figure}[t!]
  \subfigure[]{
   \label{fig:8a}
   \begin{minipage}[]{0.5\textwidth}
   \centering \includegraphics[width=2.75in]{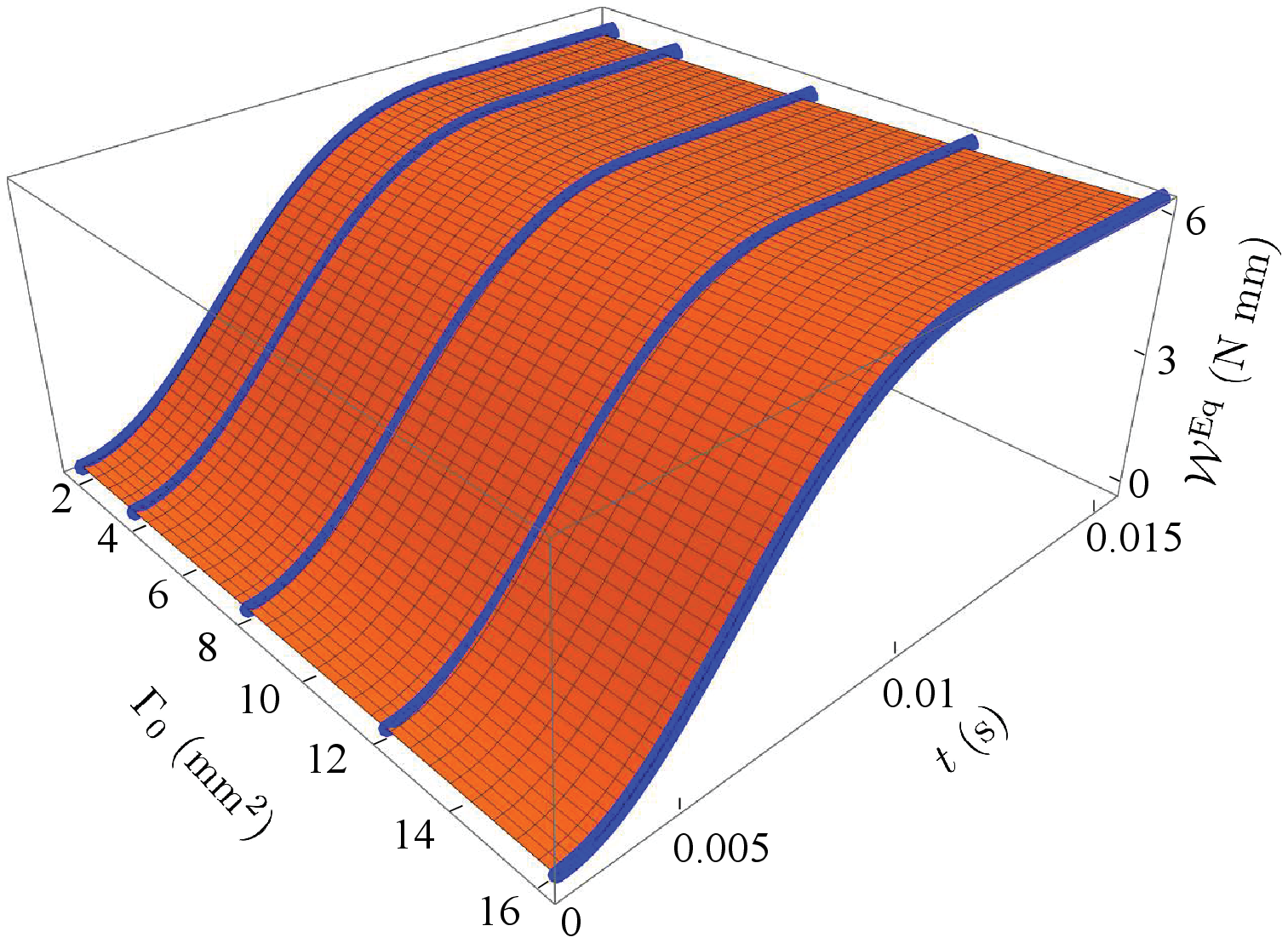}
   \vspace{0.2cm}
   \end{minipage}}
  \subfigure[]{
   \label{fig:8b}
   \begin{minipage}[]{0.5\textwidth}
   \centering \includegraphics[width=2.75in]{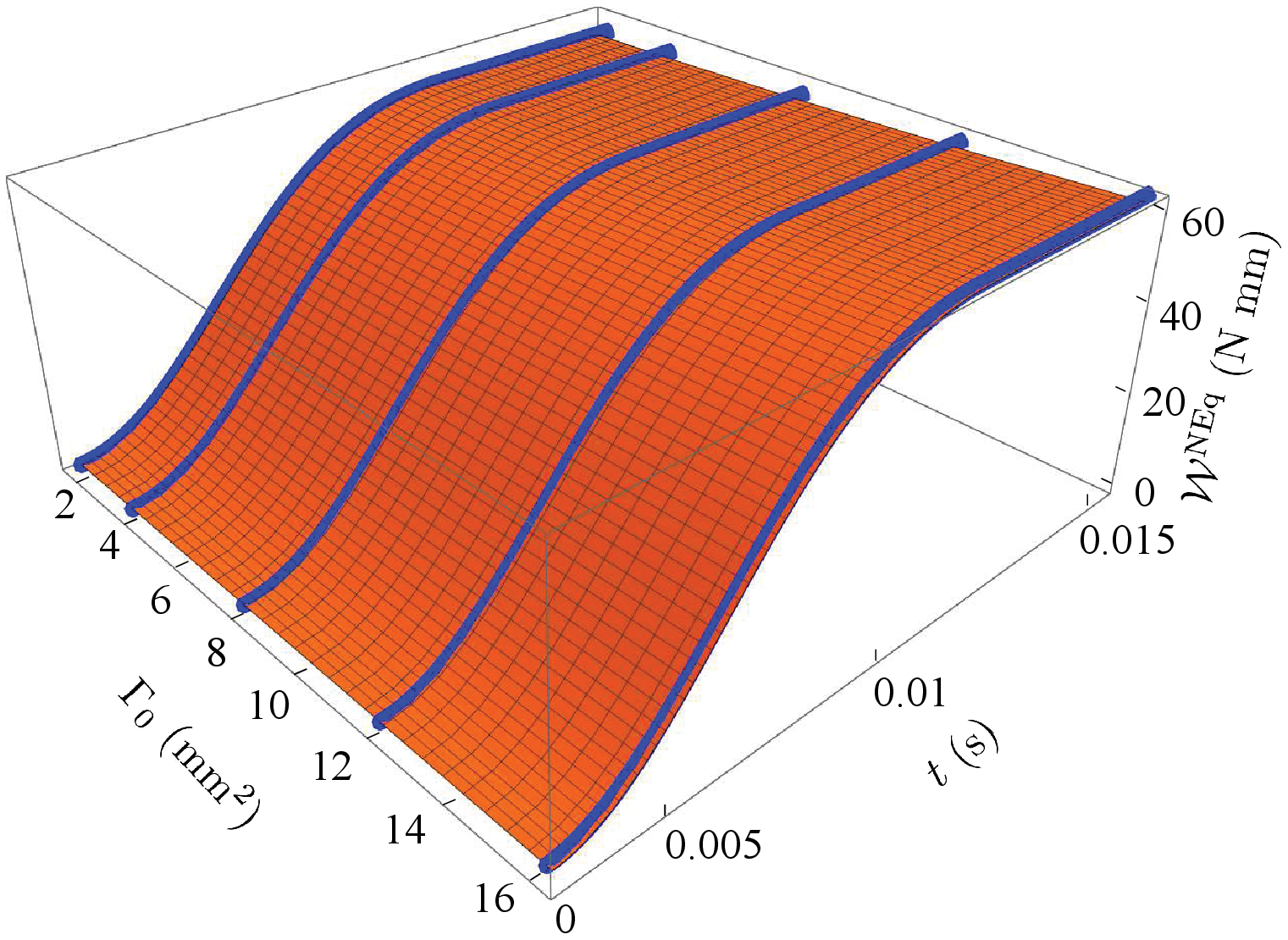}
   \vspace{0.2cm}
   \end{minipage}}
     \subfigure[]{
   \label{fig:8c}
   \begin{minipage}[]{0.5\textwidth}
   \centering \includegraphics[width=2.75in]{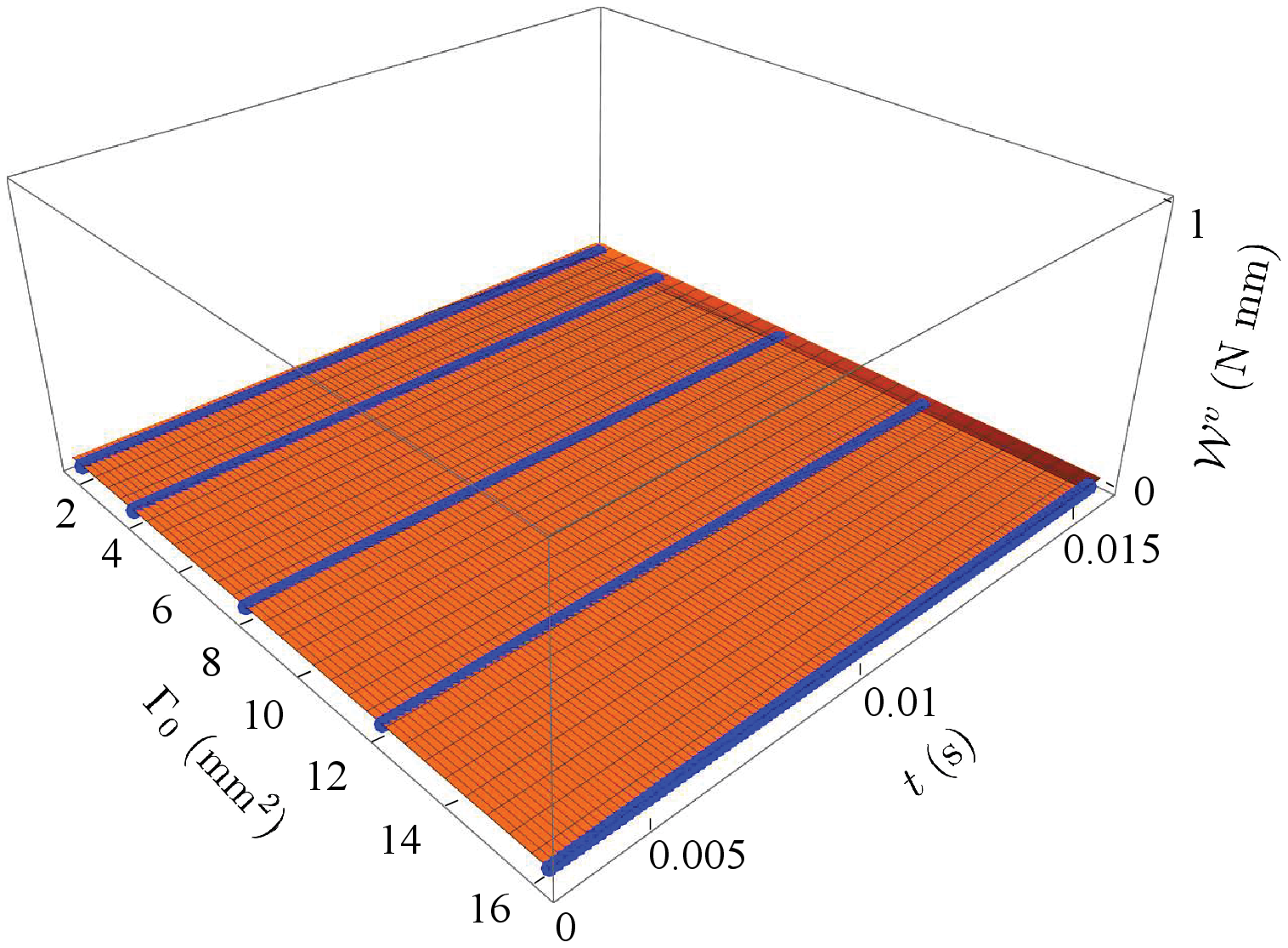}
   \vspace{0.2cm}
   \end{minipage}}
   \caption{Zoom of the time interval $t\in[0,0.015]$ s in Fig. \ref{Fig6}, focusing on the ramping of the applied force (\ref{P(t)})-(\ref{t0})$_1$ and immediately afterwards.}\label{Fig7}
\end{figure}
%

The areas under the curves in the results presented in Fig. \ref{Fig3} correspond to the total work done by the applied loads. By the same token, they correspond to the total deformation stored and dissipated by the elastomer. We thus have
\begin{equation*}
\mathcal{W}=\displaystyle\int_{H}^{h(t)} P\,{\rm d}h.
\end{equation*}
Since for this case the elastomer is a canonical elastomer with Gaussian elasticity and constant viscosity, we also have that
\begin{align}
\mathcal{W}^{{\rm Eq}}=\displaystyle\int_{\mathrm{\Omega}_0}\dfrac{\mu_1}{2}\left[{\rm tr}\,\bfC-3\right]\,{\rm d}\bfX,\label{WEq-NH}
\end{align}
\begin{align}
\mathcal{W}^{{\rm NEq}}=\displaystyle\int_{\mathrm{\Omega}_0}\dfrac{\nu_1}{2}\left[{\rm tr}(\bfC{\bfC^{v}}^{-1})-3\right]\,{\rm d}\bfX, \label{WNEq-NH}
\end{align}
and
\begin{align}
\mathcal{W}^{v}=&\mathcal{W}-\mathcal{W}^{{\rm Eq}}-\mathcal{W}^{{\rm NEq}}.  \label{Wv-NH}
\end{align}

Figures \ref{Fig6} and \ref{Fig7} show results for $\mathcal{W}^{{\rm Eq}}$, $\mathcal{W}^{{\rm NEq}}$, and $\mathcal{W}^{v}$ --- as computed from expressions (\ref{WEq-NH})-(\ref{Wv-NH}) and the pertinent numerical solutions for the deformation field $\bfy(\bfX,t)$ and the internal variable $\bfC^v(\bfX,t)$ --- for the same applied force (\ref{P(t)})-(\ref{t0})$_1$, with global stress $\sigma_0=0.3$ MPa and total time of applied loading $T=20000$ s, considered in Figs. \ref{Fig3} through \ref{Fig5}. The results are plotted as functions of the initial crack surface $\mathrm{\Gamma}_0=A \times B$ and time $t$. While Fig. \ref{Fig6} shows results for the entire duration of the loading process $t\in[0,T]$, Fig. \ref{Fig7} shows results that focus on the ramping of the applied force and immediately afterwards, over the time interval $t\in[0,0.015]$ s.

Several comments are in order. All three parts of the deformation energy appear to depend nonlinearly on both the crack surface $\mathrm{\Gamma}_0$ and time $t$. Distinctly, with respect to $t$, both the equilibrium energy $\mathcal{W}^{{\rm Eq}}$ and the non-equilibrium energy $\mathcal{W}^{{\rm NEq}}$ are seen to increase sharply, while the viscous dissipated energy $\mathcal{W}^{v}$ remains negligibly small, over the short duration of the ramping of the applied force $P(t)$ up to its final constant value $P(t)=\sigma_0(B\times L)$. Beyond the ramping process, when $t>t_0=0.01$ s, the non-equilibrium energy $\mathcal{W}^{{\rm NEq}}$ decreases monotonically in time resulting in the increase of $\mathcal{W}^{v}$ and the further increase of $\mathcal{W}^{{\rm Eq}}$. Consistent with Fig. \ref{Fig4}, the values of $\mathcal{W}^{{\rm Eq}}$, $\mathcal{W}^{{\rm NEq}}$, and $\mathcal{W}^{v}$ remain practically invariant after $t=10000$ s, since the creeping process has all but concluded by then.

\subsection{The derivative $-\partial{\mathcal{W}^{{\rm Eq}}}/\partial {\mathrm \Gamma}_0$}

The type of results presented in Figs. \ref{Fig4} and \ref{Fig6}(a) for the deformation $h(t)$ between the grips and for the equilibrium elastic energy $\mathcal{W}^{{\rm Eq}}$ can be directly used to work out the corresponding results for the equilibrium elastic force (\ref{PEquilibrium}) and, in turn, those for the complementary equilibrium elastic energy (\ref{Wstar}) in order to ultimately compute the energy release rate $-\partial{\mathcal{W}^{{\rm Eq}}}/\partial {\mathrm \Gamma}_0$ by making use of the identity (\ref{ERRP}). The relevant computations go as follows.

%
\begin{figure}[b!]
   \centering \includegraphics[width=2.75in]{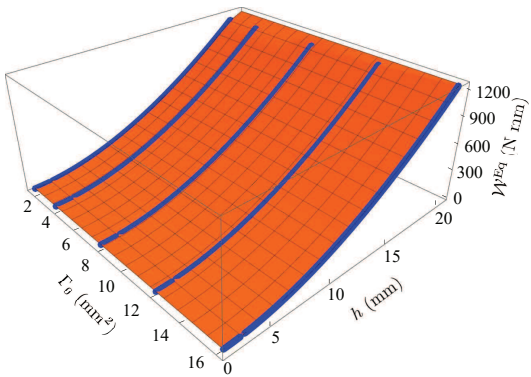}
   \caption{Equilibrium elastic energy ${\mathcal{W}^{{\rm Eq}}}$ in specimens subjected to the applied force (\ref{P(t)})-(\ref{t0})$_1$, with global stress $\sigma_0=0.3$ MPa and total time of applied loading $T=20000$ s, plotted as a function of the initial crack surface $\mathrm{\Gamma}_0=A \times B$ and the deformation $h(t)$ between the grips.}\label{Fig8}
\end{figure}
\begin{figure}[h!]
   \centering \includegraphics[width=2.75in]{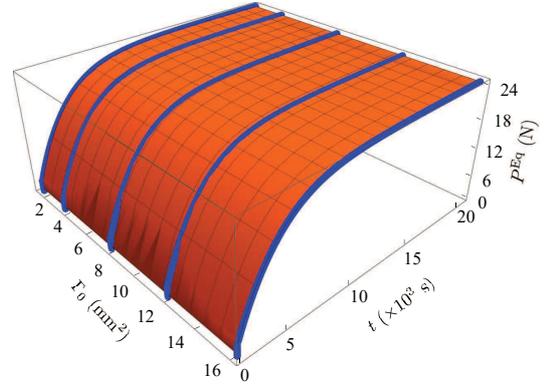}
   \caption{Equilibrium elastic force $P^{{\rm Eq}}$ in specimens subjected to the applied force (\ref{P(t)})-(\ref{t0})$_1$, with global stress $\sigma_0=0.3$ MPa and total time of applied loading $T=20000$ s, plotted as a function of the initial crack surface $\mathrm{\Gamma}_0=A \times B$ and time $t$.}\label{Fig9}
\end{figure}
\begin{figure}[h!]
   \centering \includegraphics[width=2.75in]{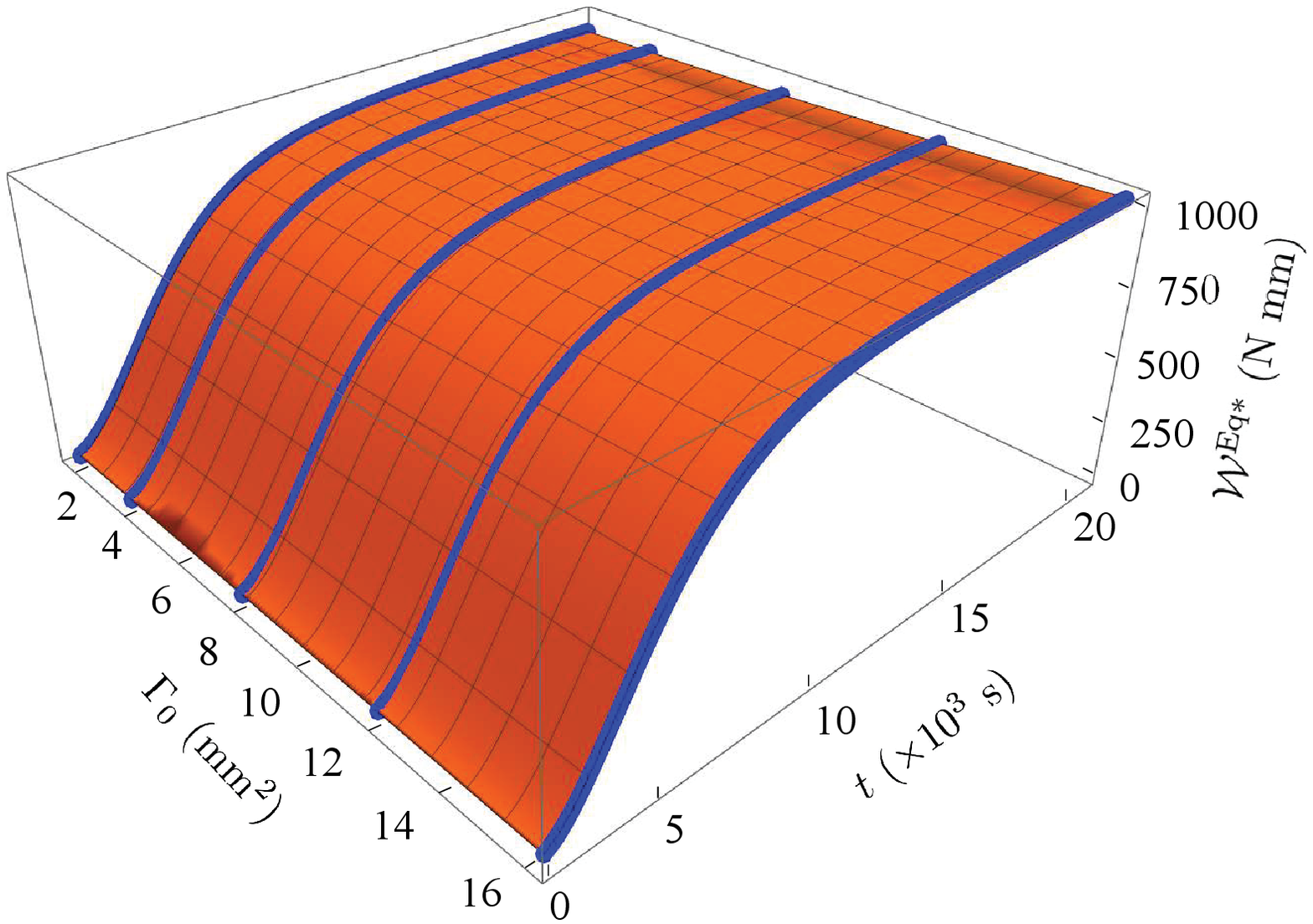}
   \caption{Complementary equilibrium elastic energy ${\mathcal{W}^{{\rm Eq}}}^\ast$ in specimens subjected to the applied force (\ref{P(t)})-(\ref{t0})$_1$, with global stress $\sigma_0=0.3$ MPa and total time of applied loading $T=20000$ s, plotted as a function of the initial crack surface $\mathrm{\Gamma}_0=A \times B$ and time $t$.}\label{Fig10}
\end{figure}

%
As a first step, for the same applied force (\ref{P(t)})-(\ref{t0})$_1$, with global stress $\sigma_0=0.3$ MPa and total time of applied loading $T=20000$ s, considered in the figures above, we replot in Fig. \ref{Fig8} the equilibrium elastic energy $\mathcal{W}^{{\rm Eq}}$, this time around,  in terms of the initial crack surface $\mathrm{\Gamma}_0=A \times B$ and the deformation $h(t)$ between the grips. From this type of 3D plot, we can readily compute the derivative (\ref{PEquilibrium}) that defines the equilibrium elastic force $P^{\rm Eq}$. The results for $P^{\rm Eq}$ from such a computation are presented in Fig. \ref{Fig9} as a function of $\mathrm{\Gamma}_0=A \times B$ and time $t$. Having determined $P^{\rm Eq}$, we can then compute the complementary equilibrium elastic energy ${\mathcal{W}^{{\rm Eq}}}^\ast$ directly from its definition (\ref{Wstar}). Figure \ref{Fig10} plots the results also as a function of $\mathrm{\Gamma}_0=A \times B$ and time $t$.
%
%
%
%

%
\begin{figure}[t!]
  \subfigure[]{
   \label{fig:10a}
   \begin{minipage}[]{0.5\textwidth}
   \centering \includegraphics[width=2.6in]{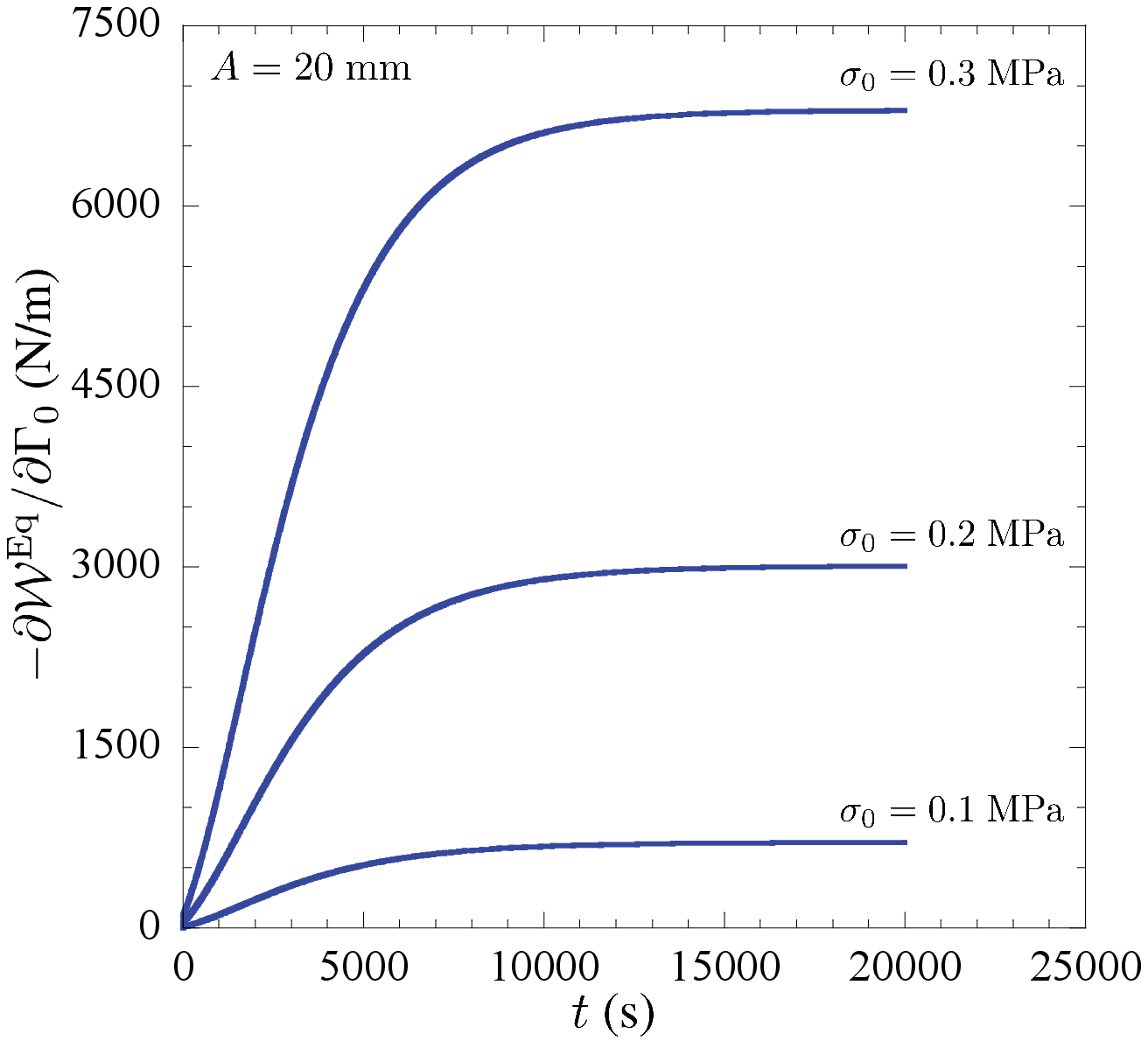}
   \vspace{0.2cm}
   \end{minipage}}
  \subfigure[]{
   \label{fig:10b}
   \begin{minipage}[]{0.5\textwidth}
   \centering \includegraphics[width=2.6in]{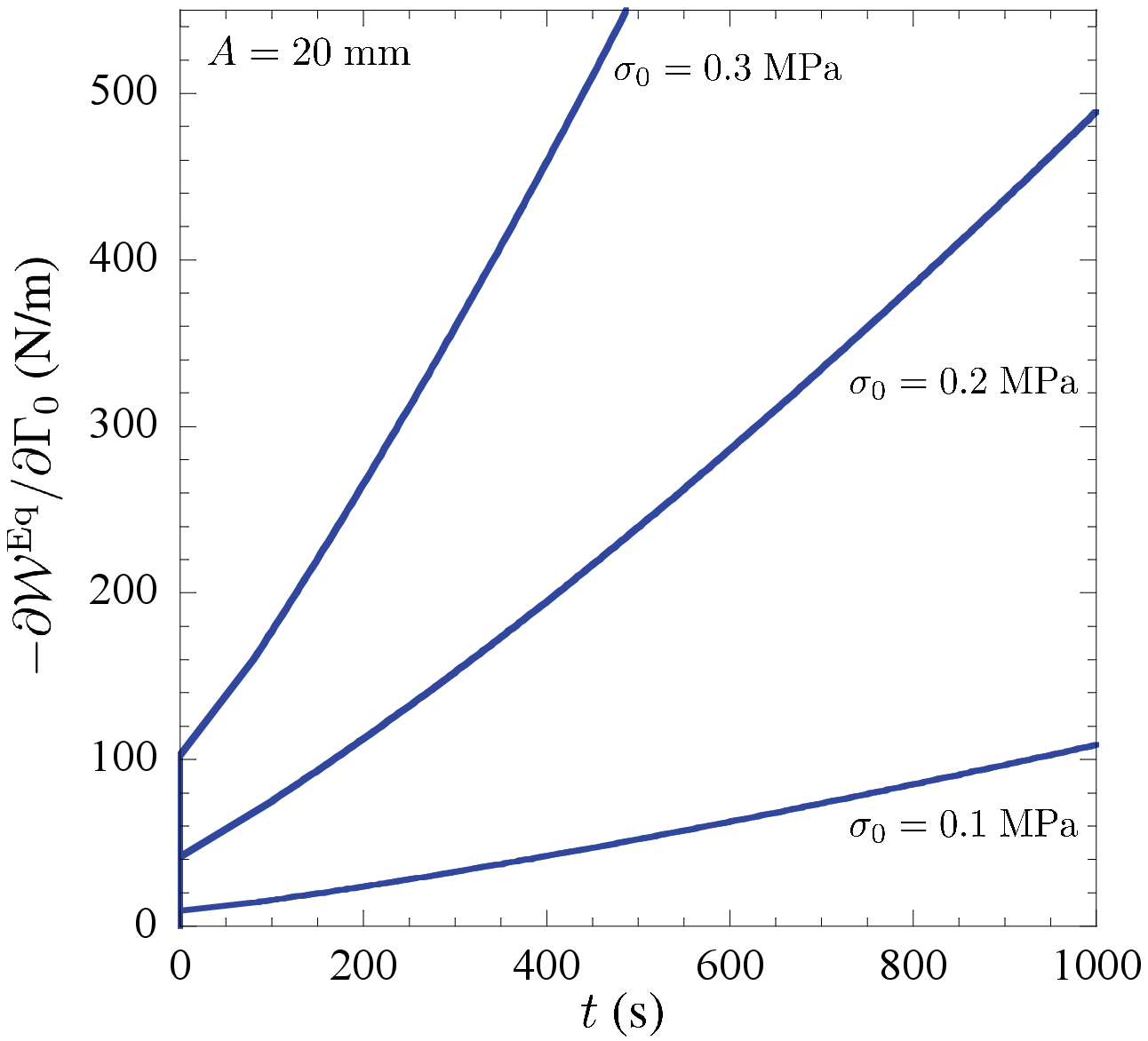}
   \vspace{0.2cm}
   \end{minipage}}
   \caption{The energy release rate $-\partial{\mathcal{W}^{{\rm Eq}}}/\partial {\mathrm \Gamma}_0$ for specimens with a pre-existing crack of length $A=20$ mm subjected to the applied force (\ref{P(t)})-(\ref{t0})$_1$ with global stresses $\sigma_0=0.1, 0.2, 0.3$ MPa and total time of applied loading $T=20000$ s, plotted as functions of time $t$. Part (a) presents results for the entire duration of the loading process $t\in[0, T]$, while part (b) zooms in the interval $t\in[0, 1000]$ s.}\label{Fig11}
\end{figure}
%

Next, from the type of 3D plots presented in Figs. \ref{Fig9} and \ref{Fig10}, we can readily compute the derivatives $\partial P^{\rm Eq}/\partial {\mathrm \Gamma}_0$ and $\partial{\mathcal{W}^{{\rm Eq}}}^\ast/\partial {\mathrm \Gamma}_0$ at fixed $P(t)$ --- which, again, it is equivalent to fixed time $t$ --- and, finally, making use of the identity (\ref{ERRP}), the energy release rate $-\partial{\mathcal{W}^{{\rm Eq}}}/\partial {\mathrm \Gamma}_0$. Figure \ref{Fig11} reports such a computation of $-\partial{\mathcal{W}^{{\rm Eq}}}/\partial {\mathrm \Gamma}_0$ for specimens with a pre-existing crack of length $A=20$ mm subjected to the global stresses $\sigma_0=0.1, 0.2, 0.3$ MPa. While part (a) of the figure shows the results as functions of time for the entire duration of the loading process $t\in[0, T]$, part (b) shows results that focus on the first 1000 s.

There are two crucial observations to be made from Fig. \ref{Fig11} that lay bare the key features of the phenomenon of delayed fracture in elastomers. First, irrespective of the applied global stress, the energy release rate $-\partial{\mathcal{W}^{{\rm Eq}}}/\partial {\mathrm \Gamma}_0$ is bounded from above and increases monotonically in time until reaching an asymptotic maximum. Second, specimens subjected to larger global stresses lead to larger values of $-\partial{\mathcal{W}^{{\rm Eq}}}/\partial {\mathrm \Gamma}_0$ at the same instance in time $t$.

According to the Griffith criticality condition (\ref{Gc-0}), the first observation entails that delayed fracture will occur --- that is, the condition $-\partial{\mathcal{W}^{{\rm Eq}}}/\partial {\mathrm \Gamma}_0$ $=G_c$ will be reached at some $t\in(t_0,T)$ --- if the applied load is between two threshold values, say $\sigma_0^{max}$ and $\sigma_0^{min}$. If the applied load is above the upper threshold $\sigma_0^{max}$, fracture will take place  during the ramping process of the load at some $t\in(0,t_0]$, without delay. If it is below the lower threshold $\sigma_0^{min}$, fracture will never occur.

On the other hand, the second observation entails that, for the same size of the pre-existing crack, specimens subjected to larger global stresses will exhibit a shorter delay for fracture to take place. The next subsection details this behavior.

\subsection{The critical time $t_c$ at fracture}

%
\begin{figure}[b!]
   \centering \includegraphics[width=2.5in]{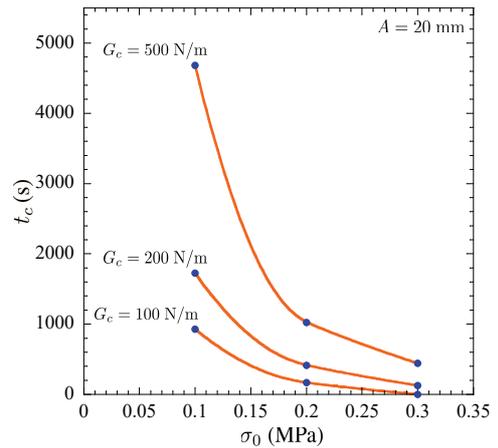}
   \caption{The critical time $t_c$ at which fracture nucleates in specimens with a pre-existing crack of length $A=20$ mm subjected to the applied force (\ref{P(t)})-(\ref{t0})$_1$. The results are shown as functions of the global stress $\sigma_0$ for three representative values of the intrinsic fracture energy $G_c$ of the elastomer.}\label{Fig12}
\end{figure}
%

Having generated the type of results presented in Fig. \ref{Fig11} for the energy release rate $-\partial{\mathcal{W}^{{\rm Eq}}}/\partial {\mathrm \Gamma}_0$ vs. time $t$ --- assuming that we also have knowledge of the intrinsic fracture energy $G_c$ of the elastomer --- we can readily determine from the Griffith criticality condition (\ref{Gc-0}) the critical time $t_c$ at which fracture will nucleate from the pre-existing crack in the specimens. This amounts to identifying the intercept of the curve $-\partial{\mathcal{W}^{{\rm Eq}}}/\partial {\mathrm \Gamma}_0$ vs. $t$ with the line $G_c$ vs. $t$.

For specimens with a pre-existing crack of length $A=20$ mm, Fig. \ref{Fig12} presents results for $t_c$ as a function of the applied global stress $\sigma_0$ for three representative values of the intrinsic fracture energy, $G_c=100, 200, 500$ N/m.

As foretold in the general conclusions established above, note that, for a given $G_c$, fracture takes shorter to nucleate in specimens subjected to larger global stresses. By the same token, for a given $\sigma_0$, fracture takes shorter to nucleate in specimens made of elastomers with smaller intrinsic fracture energies.

\section{Comparisons with the experiments of Knauss (1970) on Solithane 113}\label{Results Solithane}

We finally turn to deploying the Griffith criticality condition (\ref{Gc-0}) to explain the delayed fracture experiments of Knauss (1970) on the polyurethane elastomer Solithane 113; since the elastomer was prepared from equal amounts by volume of resin and catalyst, it is also referred to as Solithane 50/50. As noted above, these appear to be the first experiments reported in the literature that showed that elastomers can exhibit delayed fracture. The focus is on the results for specimens with the same geometry considered in the two preceding sections ($L=101.6$ mm, $H=L=101.6$ mm, $B=0.7938$ mm), featuring a pre-existing central crack of length $A=5.08$ mm, subjected to the applied global stresses\footnote{There is some uncertainty about the precise values of the global stress $\sigma_0$ applied in the experiments, since the data in Fig. 9 of (Knauss, 1970) is presented normalized by a factor ($\sigma_{{\rm g}\infty}$) that was not spelled out fully explicitly. The values $\sigma_0=0.10, 0.12, 0.13, 0.15$ MPa that we use here are our best estimate based on the information provided.} $\sigma_0=0.10, 0.12, 0.13, 0.15$ MPa at a temperature of $0$ $^{\circ}$C; see Fig. 9 in (Knauss, 1970).

\subsection{The viscoelastic behavior and intrinsic fracture energy of Solithane 113}

As emphasized in the Introduction, the use of the Griffith criticality condition (\ref{Gc-0}) requires knowledge of only two fundamental properties of the elastomer of interest: (\emph{i}) its viscoelastic behavior, from which the storage of equilibrium elastic energy can be identified, and (\emph{ii}) its intrinsic fracture energy. Both of these properties can be measured experimentally once and for all by means of conventional tests.

\subsubsection{The viscoelastic behavior}

A few years before Knauss (1970) published his findings on delayed fracture, as part of his PhD thesis work, Mueller (1968) reported a range of experimental results on the mechanical behavior of the same Solithane 113 tested by Knauss (1970). Most of these restricted attention to small deformations, but Mueller (1968) did include a handful of results involving finite deformations for the viscoelastic response of Solithane 113 under uniaxial tension applied at various constant stretch rates at a temperature of $-5$ $^{\circ}$C; see Fig. 16 in (Mueller, 1968) and also Fig. 4 in (Mueller and Knauss, 1971).

\begin{table}[b!]\centering
\caption{Values of the material constants in the viscoelastic model (\ref{S-KLP})-(\ref{Evolution-KLP}) for the polyurethane elastomer Solithane 113.}
\begin{tabular}{llll}
\hline
$\mu_1=0.2099$ MPa & & & $\mu_2=2.040\times 10^{-5}$ MPa  \\
$\alpha_1=1.941$ & & & $\alpha_2=9.344$  \\
$\nu_1=2.300$ MPa & & & $\nu_2=4.147\times10^{-2}$ MPa \\
$\beta_1=0.5353$ & & & $\beta_2=7.108$ \\
$\eta_0=150$ MPa s & & & $\eta_{\infty}=0$ MPa s \\
$K_1=2.653$ MPa s & & & $K_2=0$ MPa$^{-2}$  \\
$\gamma_1=7.977$ & & & $\gamma_2=1$ \\
\hline
\end{tabular} \label{Table1}
\end{table}

\begin{figure}[t!]
   \centering \includegraphics[width=2.4in]{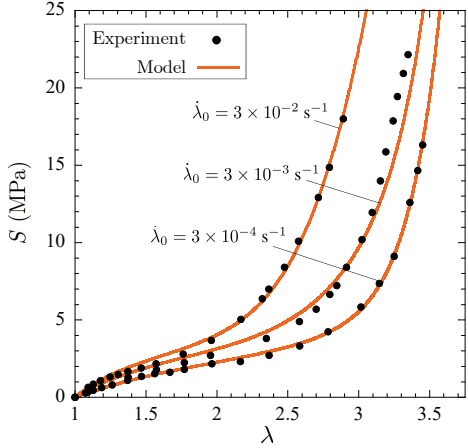}
   \caption{Comparison between the stress-stretch response (solid line) predicted by the viscoelastic model (\ref{S-KLP})-(\ref{Evolution-KLP}), with the material constants in Table \ref{Table1}, and the experimental data (solids circles) reported by Mueller (1968) for Solithane 113 subjected to uniaxial tension applied at three different constant stretch rates, $\dot{\lambda}_0= 3\times 10^{-4}, 3\times 10^{-3}, 3\times 10^{-2}$ ${\rm s}^{-1}$.}\label{Fig13}
\end{figure}

Specializing the constitutive relation (\ref{S-KLP})-(\ref{Evolution-KLP}) to such loadings --- that is, to deformation gradients of the form $\bfF={\rm diag}(\lambda,\lambda^{-1/2},\lambda^{-1/2})$ with $\lambda=1+\dot{\lambda}_{0} t$ and first Piola-Kirchhoff stresses of the form $\bfS={\rm diag}(S,0,0)$ --- and then fitting (by least squares) its material constants to the admittedly scarce experimental data of Mueller (1968) yields the values listed in Table \ref{Table1}. As seen from the comparisons presented in Fig. \ref{Fig13}, the constitutive relation (\ref{S-KLP})-(\ref{Evolution-KLP}) with such material constants describes reasonably well the viscoelastic data (solid circles) reported by Mueller (1968).

%

\begin{remark}\label{Remark4}
The material constants listed in Table \ref{Table1} indicate that, at $-5$ $^{\circ}$C, Solithane 113 is an elastomer with non-Gaussian elasticity and nonlinear viscosity. This falls squarely within the behavior of the vast majority of elastomers.
\end{remark}

\begin{remark}\label{Remark5}
In the sequel, because of the absence of experimental data at temperatures other than $-5$ $^{\circ}$C, we make use of the constitutive relation (\ref{S-KLP})-(\ref{Evolution-KLP}) with the material constants listed in Table \ref{Table1} --- which, again, strictly apply to the behavior of Solithane 113 at  $-5$ $^{\circ}$C --- to describe the viscoelastic behavior of Solithane 113 in the delayed fracture experiments of Knauss (1970) at $0$ $^{\circ}$C. This  $5$ $^{\circ}$C difference in temperature should not be taken as negligible, since the viscosity of elastomers can change rapidly near their glass transition temperature $T_g$ and the glass transition temperature for Solithane 113 happens to be about $-20$ $^{\circ}$C (Mueller and Knauss, 1971).
\end{remark}

\subsubsection{The intrinsic fracture energy}

In his PhD thesis work, Mueller (1968) also carried out experiments aimed at measuring the intrinsic fracture energy $G_c$ of Solithane 113. A summary of these was later reported in (Mueller and Knauss, 1971). The experiments consisted in carrying out ``pure-shear'' fracture tests at various constant global stretch rates in the range $[1.7\times 10^{-4},8.3\times 10^{-3}]$ s$^{-1}$ and various constant temperatures in the range $[0,50]$ $^{\circ}$C on specimens that have been swollen with the solvent Toluene. The presence of the solvent led to the minimization of viscous dissipation. From the results of such ``pure-shear'' fracture tests, it was concluded that the intrinsic fracture energy of the swollen Solithane 113 was $G_c^{sw}=28\pm7$ N/m and that this value was independent of temperature. By making use of an argument similar to that put forth by Lake and Thomas (1967), that the intrinsic fracture energy is essentially a measure of the chain-bond strength only, Mueller and Knauss (1971) then estimated that the intrinsic fracture energy of Solithane 113 in its unswollen state is
\begin{align*}
G_c=41\pm8\, {\rm N}/{\rm m},  
\end{align*}
this estimate also being independent of temperature. This value falls squarely within the range $G_c\in[10,100]$ N/m for common hydrocarbon elastomers.

\subsection{Computation of the derivative $-\partial{\mathcal{W}^{{\rm Eq}}}/\partial {\mathrm \Gamma}_0$}

%
\begin{figure}[b!]
   \centering \includegraphics[width=2.6in]{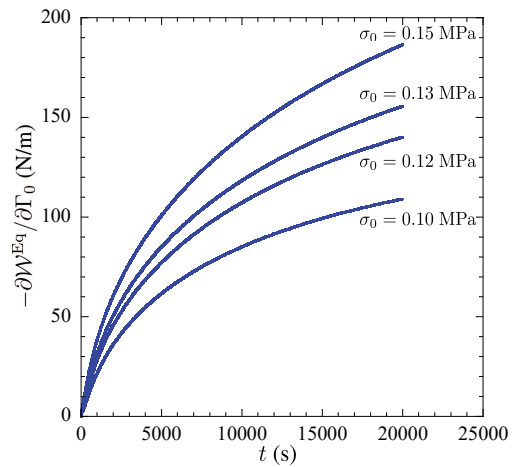}
   \caption{The energy release rate $-\partial{\mathcal{W}^{{\rm Eq}}}/\partial {\mathrm \Gamma}_0$ at $A=5.08$ mm computed from the simulations of delayed fracture tests on Solithane 113. The results correspond to applied global stresses $\sigma_0=0.10, 0.12, 0.13, 0.15$ MPa and are plotted as functions of time $t$.}\label{Fig14}
\end{figure}
%

Having established the pertinent deformation and fracture properties of Solithane 113, we proceed by repeating the same type of full-field analysis presented in Section \ref{Results Gaussian} in order to compute the derivative $-\partial\mathcal{W}^{{\rm Eq}}/\partial{\mathrm \Gamma}_0$ entering the Griffith criticality condition (\ref{Gc-0}).

Before presenting and discussing the results, the following technical remarks are in order. Since the experiments of Knauss (1970) pertain to specimens with a pre-existing crack of length $A=5.08$ mm, we perform the simulations for specimens with three crack lengths, $A=2.5, 5.08, 10$ mm. This suffices to be able to take the required derivative $-\partial{\mathcal{W}^{{\rm Eq}}}/\partial {\mathrm \Gamma}_0$ at ${\mathrm \Gamma}_0=A\times B=5.08\times 0.7938$ mm$^2$. Much like the loads used in the experiments, we carry out simulations at four different global stresses, $\sigma_0=0.10, 0.12, 0.13, 0.15$ MPa. Accordingly, in all, we carry out $3\times 4=12$ simulations of the delayed fracture tests. Furthermore, since the experiments indicate that fracture nucleates from the pre-existing crack at critical times $t_c<20000$ s, we use $T=20000$ s for the total time of applied loading in each of these simulations.

Analogous to Fig. \ref{Fig11}, Fig. \ref{Fig14} presents results for the energy release rate $-\partial{\mathcal{W}^{{\rm Eq}}}/\partial {\mathrm \Gamma}_0$ computed from the simulations of the delayed fracture tests on Solithane 113, at the applied global stresses $\sigma_0=0.10, 0.12$, $0.13$, $0.15$ MPa. Much like the results in Fig. \ref{Fig11} for the canonical case of an elastomer with Gaussian elasticity and constant viscosity, the results in Fig. \ref{Fig14} show that, irrespective of the applied global stress $\sigma_0$, the energy release rate $-\partial{\mathcal{W}^{{\rm Eq}}}/\partial {\mathrm \Gamma}_0$ increases monotonically in time towards an asymptotic maximum value. The results also show that specimens subjected to larger $\sigma_0$ lead to larger values of $-\partial{\mathcal{W}^{{\rm Eq}}}/\partial {\mathrm \Gamma}_0$ at the same instance in time $t$.

\subsection{The critical time $t_c$ at fracture}

%
\begin{figure}[b!]
   \centering \includegraphics[width=2.6in]{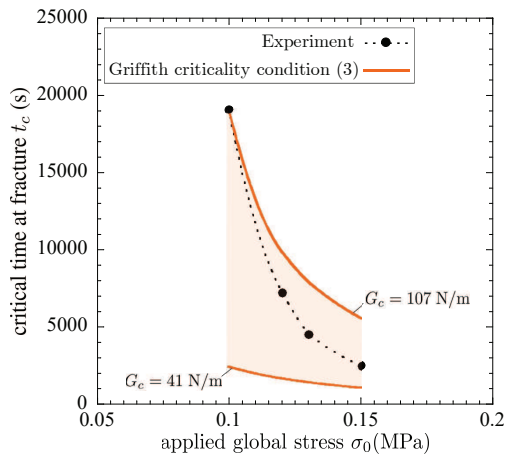}
   \caption{Comparison between the critical time $t_c$ at which fracture nucleates, according to the Griffith criticality condition (\ref{Gc-0}), and the corresponding experimental results reported by Knauss (1970) for Solithane 113 at $0$ $^{\circ}$C. The results are presented as a function of the applied global stress $\sigma_0$.}\label{Fig15}
\end{figure}
%
At this stage, we are in a position to deploy the Griffith criticality condition (\ref{Gc-0}) to explain the delayed fracture experiments of Knauss (1970).

Figure \ref{Fig15} confronts the theoretical predictions obtained from the results in  Fig. \ref{Fig14} --- specifically, again, the intercepts of the curves $-\partial{\mathcal{W}^{{\rm Eq}}}/\partial {\mathrm \Gamma}_0$ vs. $t$ with the line $G_c$ vs. $t$ --- with the corresponding experimental results (solid circles) for the critical time $t_c$ at which fracture nucleates. The results are presented as a function of the applied global stress $\sigma_0$. For the theoretical predictions, we include two results. The first one corresponds to using the average value $G_c=41$ N/m estimated by Mueller and Knauss (1971) for the intrinsic fracture energy. The second corresponds to using the somewhat larger value $G_c=107$ N/m. The experimental data falls within these two results.

Two comments are in order. First and foremost, taking into account the various sources of uncertainties (on the precise values of the applied global stress $\sigma_0$ and on the viscoelastic response of Solithane 113 at $0$ $^{\circ}$C), the Griffith criticality condition (\ref{Gc-0}) appears, indeed, to determine when delayed fracture occurs. The results in Fig. \ref{Fig15}  also make it plain that having robust experimental data for the viscoelasticity and the intrinsic fracture energy of the elastomer of interest is essential to be able to predict its delayed fracture. This is because small variations in either property may result in large changes in the critical time $t_c$ at fracture, especially when dealing with small forces that lead to long creeping processes.

\section{Final comments}\label{Sec: Final Comments}

Adding to the validation results presented by Shrimali and Lopez-Pamies (2023), who made use of the Griffith criticality condition (\ref{Gc-0}) to explain ``pure-shear'' fracture experiments carried out over a wide range of constant stretch rates on an acrylic elastomer (VHB 4905 from the company 3M), the comparisons with the delayed fracture experiments on a polyurethane elastomer presented in the preceding section provide further direct evidence that the Griffith criticality condition (\ref{Gc-0}) may indeed be the universal condition that governs crack growth in elastomers undergoing finite deformations in response to quasi-static mechanical loads.

In this context, given the recently demonstrated ability (Kumar et al. 2018a,b, 2000, 2022; Kumar and Lopez-Pamies 2020, 2021) of the phase-field theory of fracture initiated by Kumar et al. (2018a) to describe fracture nucleation and propagation in nominally elastic brittle materials at large and given the ``seamless'' mathematical generalization that the Griffith criticality condition (\ref{Gc-0}) provides of the classical Griffith criticality for elastic brittle materials, a next sensible step would be to successively follow in the footsteps of Francfort and Marigo (1998), Bourdin et al. (2000), and Kumar et al. (2018a) in order to:
\begin{enumerate}

\item{turn the  Griffith criticality condition (\ref{Gc-0}) into a complete mathematical description of fracture nucleation from pre-existing cracks and of fracture propagation in viscoelastic elastomers,}

\item{regularize such a description into numerically tractable phase-field-type PDEs (partial differential equations), and}

\item{generalize those PDEs to account for nucleation of fracture at large (not just from large pre-existing cracks, but also from the bulk, smooth and non-smooth boundary points, and small pre-existing cracks)}

\end{enumerate}
so as to formulate a complete and numerically tractable mathematical description of the nucleation and propagation of fracture in viscoelastic materials subjected to arbitrary quasi-static mechanical loads.

\section*{Acknowledgements}

\noindent Support for this work by the National Science Foundation through the Grants CMMI-1901583 and CMMI-2132528 is gratefully acknowledged.


\begin{thebibliography}{99}

\bibitem{Ahagon-Gent75}
Ahagon A, Gent AN (1975). Threshold fracture energies for elastomers. J Polym Sci Polym Phys 13:1903--1911.

\bibitem{Bourdin00}
Bourdin B, Francfort GA,  Marigo JJ (2000). Numerical experiments in revisited brittle fracture. Journal of the Mechanics and Physics of Solids 48:797--826.

\bibitem{Ravi2022}
Chen S, Ravi-Chandar K (2022). Nonlinear poroviscoelastic behavior of gelatin-based hydrogel. Journal of the Mechanics and Physics of Solids 158:104650.

\bibitem{Cohen21}
Chockalingam S, Roth C, Henzel T, Cohen T (2021). Probing local nonlinear viscoelastic properties in soft materials. Journal of the Mechanics and Physics of Solids 146:104172.

\bibitem{Christensen79}
Christensen RM (1979). A rate-dependent criterion for crack growth. Int J Fract 15:3--21.

\bibitem{deGennes96}
de Gennes PG (1996). Soft adhesives. Langmuir 12:4497--4500.

\bibitem{FrancfortMarigo98}
Francfort GA,  Marigo JJ (1998). Revisiting brittle fracture as an energy minimization problem. Journal of the Mechanics and Physics of Solids 46:1319--1342.

\bibitem{Gent96}
Gent AN (1996). Adhesion and strength of viscoelastic solids. Is there a relationship between adhesion and bulk properties?. Langmuir 12:4492--4496.

\bibitem{Gent82}
Gent AN, Tobias RH (1982). Threshold tear strength of elastomers. J. Polym. Sci. Polym. Phys. 20:2051--2058.

\bibitem{GentLai94}
Gent AN, Lai SM (1994). Interfacial bonding, energy dissipation, and adhesion. Journal of Polymer Science: Part B: Polymer Physics 32:1543--1555.

\bibitem{GLP21}
Ghosh K, Lopez-Pamies O (2021). On the two-potential constitutive modeling of dielectric elastomers. Meccanica 56:1505--1521.

\bibitem{GSKLP21}
Ghosh K, Shrimali B, Kumar A, Lopez-Pamies O (2021). The nonlinear viscoelastic response of suspensions of rigid inclusions in rubber: I --- Gaussian rubber with constant viscosity. Journal of the Mechanics and Physics of Solids 154:104544.

\bibitem{Greensmith55}
Greensmith HW, Thomas AG (1955). Rupture of rubber. III. Determination of tear properties. Journal of Polymer Science 18:189--200.


\bibitem{Knauss70}
Knauss WG (1970). Delayed failure --- the Griffith problem for linearly viscoelastic materials. Int J Fract Mech 6:7--20.

\bibitem{Knauss1973}
Knauss WG (1973). On the steady propagation of a crack in a viscoelastic sheet: Experiments and analysis. In proceedings of the Deformation and Fracture of
High Polymers, pp 501--541.

\bibitem{Knauss15}
Knauss WG (2015). A review of fracture in viscoelastic materials. Int J Fract 196: 99--146.

\bibitem{KLP16}
Kumar A, Lopez-Pamies O (2016). On the two-potential constitutive modelling of rubber viscoelastic materials. Comptes Rendus Mecanique 344:102--112.

\bibitem{KFLP18a}
Kumar A, Francfort GA, Lopez-Pamies O (2018a). Fracture and healing of elastomers: A phase-transition theory and numerical implementation. Journal of the Mechanics and Physics of Solids 112:523--551.

\bibitem{KRLP18b}
Kumar A, Ravi-Chandar K, Lopez-Pamies O (2018b). The configurational-forces view of fracture and healing in elastomers as a phase transition. Int J Fract 213:1--16.

\bibitem{KLP20}
Kumar A, Lopez-Pamies O (2020). The phase-field approach to self-healable fracture of elastomers: A model accounting for fracture nucleation at large, with application to a class of conspicuous experiments. Theoret Appl Fract Mech 107:102550.

\bibitem{KBFLP20}
Kumar A, Bourdin B, Francfort GA, Lopez-Pamies O (2020). Revisiting nucleation in the phase-field approach to brittle fracture. Journal of the Mechanics and Physics of Solids 142:104027.

\bibitem{KLP21}
Kumar A, Lopez-Pamies O (2021). The poker-chip experiments of Gent and Lindley (1959) explained. J Mech Phys Solids 150:104359.

\bibitem{KRLP22}
Kumar A, Ravi-Chandar K, Lopez-Pamies O (2022). The revisited phase-field approach to brittle fracture: Application to indentation and notch problems. Int J Fract 237:83--100.

\bibitem{LakeandThomas67}
Lake GJ, Thomas AG (1967). The strength of highly elastic materials. Proceedings of the Royal Society of London A 300:108--119.


\bibitem{LP10}
Lopez-Pamies O (2010). A new $I_1$-based hyperelastic model for rubber elastic materials. Comptes Rendus Mecanique 338:3--11.

\bibitem{Mueller1968}
Mueller HK (1968). Stable Crack Propagation in a Viscoelastic Strip. Ph.D. Dissertation. California Institute of Technology.

\bibitem{Knauss71}
Mueller HK, Knauss WG (1971). The fracture energy and some mechanical properties of a polyurethane elastomer. Trans Soc Rheo 15:217--233.

\bibitem{Mullins59}
Mullins L (1959). Rupture of rubber. IX. Role of hysteresis in the tearing of rubber. Transactions of the Institution of the Rubber Industry 35:213--222.

\bibitem{Persson05}
Persson BNJ, Brener EA (2005). Crack propagation in viscoelastic solids. Phys Rev E 71:036123.

\bibitem{Rivlin1953}
Rivlin RS, Thomas AG (1953). Rupture of rubber. I. Characteristic energy for tearing. Journal of Polymer Science 10:291--318.

\bibitem{Schapery1975}
Schapery  RA (1975). A theory of crack initiation and growth in viscoelastic media --- I. Theoretical development. Int J Fract 11:141--159.

\bibitem{Schapery1984}
Schapery RA (1984). Correspondence principles and a generalized $J$ integral for large deformation and fracture analysis of viscoelastic media. Int J Fract 25:195--223.

\bibitem{SLP23}
Shrimali B, Lopez-Pamies O (2023). The ``pure-shear'' fracture test for viscoelastic elastomers and its revelation on Griffith fracture. Extreme Mechanics Letters 58:101944.


\bibitem{Thomas00}
Tsunoda K, Busfield JJC, Davies CKL, Thomas AG (2000). Effect of materials variables on the tear behaviour of a non-crystallising elastomer. Journal of Materials Science 35:5187--5198.

\bibitem{Zener48}
Zener CM (1948). Elasticity and anelasticity of metals. University of Chicago Press, Chicago.











\end{thebibliography}
\end{document}